\documentclass[pdflatex,sn-nature]{sn-jnl}
\usepackage{amsmath,amssymb,graphicx,booktabs}
\graphicspath{{figures/}}
\setcitestyle{super,open={},close={}}

\begin{document}
\title{All-electron machine-learned orbital-free density functional for periodic CHNO systems}

\author[1]{\fnm{Marc K.} \sur{Ickler}}
\author[2]{\fnm{Andreas} \sur{Hermann}}
\author*[1]{\fnm{Fred A.} \sur{Hamprecht}}\email{fred.hamprecht@iwr.uni-heidelberg.de}
\affil[1]{\orgdiv{Interdisciplinary Center for Scientific Computing (IWR)},
\orgname{Heidelberg University}, \orgaddress{\city{Heidelberg},
\postcode{69120}, \country{Germany}}}
\affil[2]{\orgdiv{Centre for Science at Extreme Conditions and SUPA,
School of Physics and Astronomy}, \orgname{University of Edinburgh},
\orgaddress{\city{Edinburgh}, \postcode{EH9 3FD}, \country{United Kingdom}}}

\abstract{We introduce STRUCTURES-P, a machine-learned all-electron
orbital-free density functional for periodic CHNO systems. Electrostatics are
evaluated analytically in an atom-centered Gaussian basis, while an equivariant
network learns the kinetic and exchange--correlation energy from energies and
gradients of non-ground-state densities. Minimizing the functional from a
superposition of atomic densities recovers the Kohn--Sham ground state: on
held-out crystals, the density error falls from 6.7\% to 0.4\% of the electron
count and the energy error reaches 0.25~mHa per atom, half that of a
same-architecture interatomic potential. A single functional can be trained for
both isolated molecules and periodic systems, and the model generalizes to
larger unit cells. It largely preserves relative-energy rankings and reproduces
convex-hull vertices on the tested H--N composition slice. Together, these
results demonstrate the viability of machine-learned all-electron orbital-free
DFT for variational density optimization and energy evaluation in
fixed-geometry solids.}
\maketitle

\section{Introduction}
Kohn--Sham density functional theory
(KS-DFT)\cite{kohnSelfConsistentEquationsIncluding1965} is central to
predicting material properties, but its reliance on orbitals makes large
candidate-structure searches computationally expensive.
Orbital-free DFT (OF-DFT)\cite{hohenbergInhomogeneousElectronGas1964} instead
minimizes an explicit density functional for the total energy over the
electron density itself, without introducing orbitals, yielding the
ground-state density as the minimizer and the ground-state energy as the
minimum. We refer to the
iterative procedure for this variational minimization as density optimization. By avoiding the
repeated diagonalization of the Kohn--Sham Hamiltonian, whose cost scales
nominally as the cube of the system size, OF-DFT opens a route to lower-scaling
electronic-structure calculations\cite{miOrbitalFreeDensityFunctional2023b}.
Its broader application, however, remains limited by the lack of transferable
approximations to the noninteracting kinetic energy, particularly for
covalently bonded systems and for the non-ground-state densities encountered
during variational optimization\cite{xiaCanOrbitalfreeDensity2012}.
Here we show that a machine-learned functional can be variationally minimized
over the explicit all-electron density of three-dimensional periodic crystals.
Our model, STRUCTURES-P, combines analytically evaluated periodic
electrostatics with a learned kinetic plus exchange--correlation energy, and
reaches sub-millihartree-per-atom energy errors on held-out CHNO crystals.

Conventional kinetic-energy density functionals build on the local
Thomas--Fermi and gradient-dependent von Weizs\"acker
approximations\cite{thomasCalculationAtomicFields1927,fermiStatistischeMethodeZur1928,weizsaeckerZurTheorieKernmassen1935}.
The nonlocal Wang--Teter and Wang--Govind--Carter functionals use the linear
response of the uniform electron gas and are successful particularly for
nearly-free-electron metals\cite{wangKineticenergyFunctionalElectron1992,wangOrbitalfreeKineticenergyDensity1999}.
The Huang--Carter functional extends this approach to semiconductors and
covalent solids\cite{huangNonlocalOrbitalfreeKinetic2010}, while Laplacian-level
semilocal functionals provide another route beyond generalized-gradient
approximations\cite{constantinPerformanceSemilocalKinetic2019}.
These functionals have enabled random structure searching for elemental
metals\cite{wittRandomStructureSearching2021}.
However, accuracy remains strongly dependent on material class and bonding
environment, and molecular bonding remains difficult to describe even at
equilibrium geometries\cite{xiaCanOrbitalfreeDensity2012,borgooMolecularBindingPostKohnSham2014}.
Recent inversion methods recover kinetic and Pauli potentials for prescribed
periodic densities, with demonstrations in one-dimensional systems, but do not
supply a transferable explicit
functional\cite{majumdarOrbitalFreeDensityFunctional2025}.

Machine learning offers flexible approximations to the nonlocal density
dependence of the kinetic-energy functional\cite{chenMachineLearningEnhancedOrbitalFree2026}.
Early work learned kinetic-energy functionals in model systems or on molecular
densities\cite{snyderFindingDensityFunctionals2012a,yaoKineticEnergyHydrocarbons2016a,seinoSemilocalMachinelearnedKinetic2018a},
but also exposed a central difficulty: a small energy interpolation error does
not imply a sufficiently accurate functional derivative for variational
optimization. Subsequent approaches trained energies and functional derivatives
jointly\cite{meyerMachineLearningApproaches2020a}, learned separate semilocal
functionals and potentials for Gaussian-represented molecular
densities\cite{fujinamiOrbitalfreeDensityFunctional2020}, explored direct
minimization for periodic graphene\cite{ryczkoOrbitalFreeDensityFunctional2022a},
or imposed the variational principle as a
regularizer\cite{mazo-sevillanoVariationalPrincipleRegularize2023}.
A semilocal kinetic-energy functional trained on Kohn--Sham functional derivatives also enabled
self-consistent OF-DFT for atoms, diatomic molecules, and periodic solids with
local pseudopotentials\cite{imotoOrderOrbitalfreeDensityfunctional2021}.
The relevant test is thus not just how accurately a learned functional interpolates
energies on reference densities, but whether it can be minimized from an
initial density to recover a meaningful density and energy.

Sun and Chen demonstrated self-consistent density optimization with a learned
nonlocal kinetic-energy functional for Li, Mg, Al and their alloys using local
pseudopotentials\cite{sunMachineLearningBased2024}. A separately trained
multichannel extension was evaluated on silicon and III--V
semiconductors\cite{sunMultichannelMachineLearning2024}.
Other periodic studies evaluated learned kinetic energies on Kohn--Sham
reference densities or as non-self-consistent
corrections\cite{manzhosDatadrivenKineticEnergy2020,luderMachinelearnedKineticEnergy2024,thapaOrbitalfreeDensityFunctionals2025};
an ML-guided analytical functional also demonstrated self-consistent density
optimization for Al and Si\cite{manzhosMachineLearningguidedConstruction2025a}.

Periodic OF-DFT implementations generally employ local pseudopotentials, which
replace the nuclear potential and core electrons with a smoother effective
interaction. Modern transferable pseudopotentials are usually nonlocal
operators\cite{hamannOptimizedNormconservingVanderbilt2013,vansettenPseudoDojoTrainingGrading2018};
their energies depend on the one-particle density matrix rather than the
density alone. Local pseudopotentials can be accurate in their target
environments but are harder to make broadly
transferable\cite{zhouTransferableLocalPseudopotentials2004,huangTransferableLocalPseudopotentials2008,delrioGloballyOptimizedLocalPseudopotentials2017,riosvargasPseudopotentialsOrbitalFree2026}.
Density-based reconstructions of the density matrix permit nonlocal
pseudopotentials at the cost of an additional approximation to the
electron--ion
energy\cite{xuNonlocalPseudopotentialEnergy2022,maNonlocalPseudopotentialEnergy2024}.
The choice of pseudopotential also changes the short-wavelength kinetic-energy
response, coupling errors in the pseudopotential and the kinetic-energy
functional\cite{moldabekovNonlocalVsLocal2025}.
An all-electron formulation avoids pseudopotentials but must resolve the rapidly
varying core density and evaluate periodic electrostatics. Periodic
all-electron OF-DFT has been implemented using projector augmented
waves\cite{lehtomakiOrbitalfreeDensityFunctional2014}, while adaptive finite-element
methods have demonstrated direct treatment of singular nuclear potentials in
atomic calculations\cite{motamarriHigherorderAdaptiveFiniteelement2012}.
All-electron Gaussian implementations with analytical functionals exist for
finite
systems\cite{chanThomasFermiDirac2001,ryleyRobustAllElectronOptimization2021,lemaitreGaussianBasisFunctions2023},
including recent work recovering near-exact nuclear
cusps\cite{hutcheonExactNuclearCusps2024}.

Machine learning has also been used to bypass the energy functional and
predict densities directly. ELECTRAFI constructs
periodic charge densities from local anisotropic
Gaussians\cite{elsborgGlobalPlaneWaves2026a};
DeePAW predicts crystal charge densities and formation energies
directly\cite{suDeePAW2026}.
Surrogate functionals occupy a middle ground: they are trained using only
ground-state data, and their objective is to recover the ground-state
density\cite{remmeSurrogateFunctionalsMachinelearned2026} and
energy\cite{wagnerOrbitalFreeSurrogateFunctionals2026} through a prescribed
optimization procedure, without requiring the functional to reproduce physical
energies away from the ground state.

The present work builds directly on all-electron learned functionals for
molecules. KineticNet introduced nonlocal equivariant learning and
training densities generated under perturbed external
potentials\cite{remmeKineticNetDeepLearning2023a}. M-OFDFT combined an
atom-centered Gaussian density representation with a nonlocal learned functional
and demonstrated accuracy and size extrapolation for
molecules\cite{zhangOvercomingBarrierOrbitalfree2024a}. STRUCTURES25 used
Tensor Frames for rotationally equivariant message
passing\cite{lippmannCanonicalizationHowTensorial2025} and perturbed-potential
training data to achieve stable variational density optimization on QM9 and
QMugs\cite{remmeStableAccurateOrbitalFree2025}.

Taken together, prior work has established self-consistent learned OF-DFT for
periodic pseudopotential systems and variational all-electron learned
functionals for finite molecules. Generalizing the latter to three-dimensional
periodic crystals, with minimization over the explicit all-electron density, has
not yet been achieved.

STRUCTURES-P is a machine-learned OF-DFT framework based on a Gaussian
density representation shared between isolated and periodic systems; here we
develop its all-electron periodic formulation for CHNO crystals.
The Hartree, electron--nuclear, and nucleus--nucleus contributions are evaluated
analytically under periodic boundary conditions, while an equivariant graph
neural network represents the noninteracting kinetic plus exchange--correlation
energy. The functional is trained on non-ground-state energies and
density-coefficient gradients and minimized at fixed electron number
(Fig.~\ref{fig:pipeline}).
Evaluated against all-electron PySCF references for held-out crystals, the
functional generalizes to larger unit cells and can be jointly trained on
isolated molecules and periodic crystals. Relative-enthalpy rankings and an
H--N convex-hull comparison test its use for fixed-geometry candidate evaluation.
To our knowledge, these results provide the first demonstration of learned
OF-DFT enabling variational optimization of the explicit all-electron density
under three-dimensional periodic boundary conditions.

\section{Results}
\subsection{A variational all-electron functional for periodic crystals}
\label{sec:framework}
\begin{figure}[htbp]
    \centering
    \includegraphics[width=\textwidth]{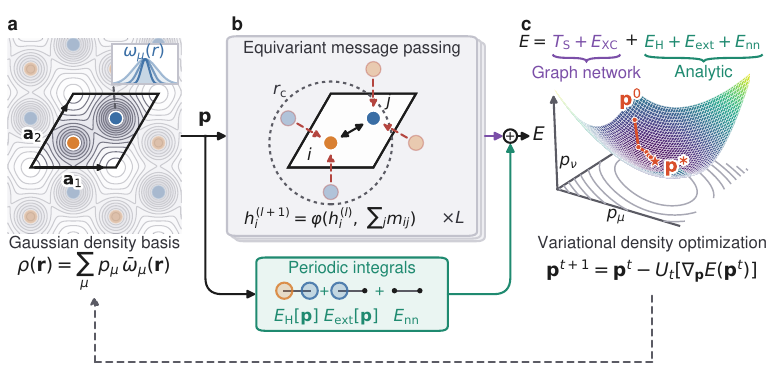}
    \caption{Overview of STRUCTURES-P for periodic crystals with Gaussian-type density
    basis functions, drawn here for a two-dimensional lattice.
    \textbf{a}~The electron density of the primitive cell is represented by
    coefficients $\mathbf{p}$ of atom-centered Gaussians,
    $\rho(\mathbf{r})=\sum_\mu p_\mu\bar\omega_\mu(\mathbf{r})$, where
    $\bar\omega_\mu$ includes the lattice images; contours show the
    resulting periodic density, the inset the radial profiles of the
    even-tempered shells.
    \textbf{b}~The same structure defines a periodic graph, on which each atom of the
    cell receives messages from its neighbors within a cutoff
    $r_\mathrm{c}$, including periodic images. An equivariant graph network
    (eSEN) predicts the kinetic and exchange--correlation energy $T_\mathrm{S}+E_\mathrm{XC}$, while
    the Hartree, nuclear-attraction, and nucleus--nucleus energies are evaluated
    analytically from periodic two-center integrals; of these, only
    $E_\mathrm{H}$ and $E_\mathrm{ext}$ depend on $\mathbf{p}$.
    \textbf{c}~Together these terms define the total energy, whose gradient with
    respect to $\mathbf{p}$ (by automatic differentiation) drives a variational
    density optimization from an initial guess $\mathbf{p}^0$ to an
    optimized density $\mathbf{p}^\ast$ under a fixed-electron-number constraint;
    the update step $U_t$ is specified in Section~\ref{sec:density_optimization}.}
    \label{fig:pipeline}
\end{figure}

STRUCTURES-P extends the molecular STRUCTURES25
functional\cite{remmeStableAccurateOrbitalFree2025} to periodic all-electron
systems. It keeps the Gaussian density representation and the variational
training strategy, and adds analytically evaluated periodic electrostatics, a
periodic label pipeline, an equivariant network with periodic message
passing, and a regularized density optimization for crystals.
Figure~\ref{fig:pipeline} summarizes the inference path: represent the density,
evaluate the total energy, and optimize the density coefficients at fixed
crystal geometry. We describe this path first, followed by the training data
that make the optimization possible.

We expand the density in atom-centered basis functions, chosen here to be
Gaussians. For a periodic crystal, each density basis function is repeated over
the Bravais lattice $\mathcal{L}$ of the unit cell $\Omega$,
\begin{equation}
    \rho(\mathbf{r}) =
    \sum_{\mu=1}^{N_\mathrm{OF}} p_\mu
    \sum_{\mathbf{T}\in\mathcal{L}}
    \omega_\mu(\mathbf{r}-\mathbf{R}_{a(\mu)}-\mathbf{T}),
    \label{eq:density}
\end{equation}
where $\mu$ runs over the $N_\mathrm{OF}$ basis functions associated with atoms
in one unit cell and $a(\mu)$ denotes the atom carrying basis function $\mu$.
The same coefficient vector $\mathbf{p}$ is shared by all periodic images.
We write $\bar\omega_\mu(\mathbf{r})$ for the lattice sum in
Eq.~\eqref{eq:density}. For a fixed density basis, the density is fully specified
by $\mathbf{p}$ and the crystal geometry
$\mathcal{M}=\{\mathbf{R}_a,Z_a,\mathbf{A}\}$, where $\mathbf{R}_a$ and $Z_a$
are the atomic positions and atomic numbers, and $\mathbf{A}$ contains the cell
vectors. The basis represents the density, not orbitals. For isolated systems,
the expansion reduces to the $\mathbf{T}=\mathbf{0}$ term of the lattice sum in
Eq.~\eqref{eq:density}. This shared atom-centered density representation allows
a jointly trained functional to use one set of parameters for molecules and
periodic crystals.

An equivariant graph neural network represents the kinetic and
exchange--correlation contribution,
$F_\theta(\mathbf{p},\mathcal{M})\approx T_\mathrm{S}[\rho]+E_\mathrm{XC}[\rho]$.
Here, $T_\mathrm{S}$ is the noninteracting kinetic energy. In the Kohn--Sham
partition, $E_\mathrm{XC}$ contains both the interacting kinetic correction
$T-T_\mathrm{S}$ and the nonclassical electron--electron interaction.
Conventional OF-DFT approximates $T_\mathrm{S}$ and $E_\mathrm{XC}$
separately. To avoid explicit orbital dependence, conventional implementations
typically use local or semilocal exchange--correlation approximations rather
than orbital-dependent hybrids. This separation is not a formal requirement
of OF-DFT; here, we learn $T_\mathrm{S}+E_\mathrm{XC}$ jointly.
The Hartree energy $E_\mathrm{H}$, electron--nuclear energy
$E_\mathrm{ext}$, and nucleus--nucleus energy $E_\mathrm{nn}$ are evaluated
analytically, giving the total energy
\begin{equation}
    E_\theta(\mathbf{p},\mathcal{M})
    =
    F_\theta(\mathbf{p},\mathcal{M})
    + E_\mathrm{H}(\mathbf{p})
    + E_\mathrm{ext}(\mathbf{p}, \mathcal{M})
    + E_\mathrm{nn}(\mathcal{M}).
    \label{eq:model_total_energy}
\end{equation}

The linear density expansion makes the Hartree energy quadratic in
$\mathbf{p}$ and the nuclear-attraction energy linear; the nuclear repulsion
is independent of $\mathbf{p}$. The matrices and vectors defining the Hartree
and nuclear-attraction terms depend only on geometry and are computed once per
structure, then reused throughout density optimization.
Together with the learned contribution, this permits density optimization
without building or diagonalizing a Kohn--Sham Hamiltonian.

At inference time, we minimize this total energy with respect to the density
coefficients under a fixed-electron-number constraint,
\begin{equation}
    \mathbf{p}^\star
    =
    \arg\min_{\mathbf{p}} E_\theta(\mathbf{p},\mathcal{M})
    \quad
    \mathrm{s.t.}\quad
    \mathbf{q}^{\mathrm{T}}\mathbf{p}=N_e .
    \label{eq:density_optimization_problem}
\end{equation}
Here $q_\mu=\int_\Omega\bar\omega_\mu(\mathbf{r})\,d\mathbf{r}$ and
$N_e=\sum_a Z_a$, so the constraint keeps the unit cell neutral.
Starting from a superposition-of-atomic-densities guess, we use regularized,
charge-conserving coefficient updates driven by the energy gradient.
The optimizer and its numerical convergence criterion are specified in
Section~\ref{sec:density_optimization}; the reported densities are endpoints
of this regularized optimization, whose convergence criterion does not
establish stationarity in every charge-conserving direction.

During optimization, intermediate densities generally differ from the
Kohn--Sham ground-state density. Training therefore uses non-ground-state
samples generated by applying local-potential perturbations to converged
Kohn--Sham Fock matrices. The sampled densities are fitted into the Gaussian
density basis and supply energy and density-coefficient gradient labels for
the learned contribution. Basis construction, periodic integrals, and label
generation are detailed in Secs.~\ref{sec:density-basis}--\ref{sec:datagen},
including the approximations in the gradient labels.

We first test whether density optimization from an atomic-density
initialization recovers accurate energies and densities. We then assess whether
one jointly trained functional can describe both isolated molecules and periodic
crystals, followed by transfer to larger cells. As a complementary test, we
assess how accurately the functional ranks candidate crystal structures using
relative enthalpies at fixed geometries.

\subsection{Evaluation setup}
\begin{figure}[htbp]
    \centering
    \includegraphics[width=\textwidth]{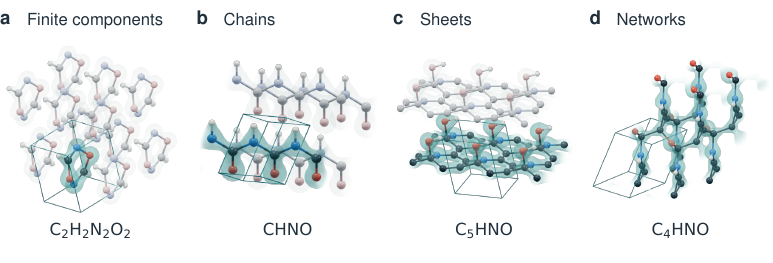}
    \caption{Selected test-set crystals illustrating the connectivity classes
    of the dataset: \textbf{a} finite bonded components (0D), \textbf{b} chains
    (1D), \textbf{c} sheets (2D), and \textbf{d} three-dimensional networks
    (3D). Surfaces show fitted Kohn--Sham reference densities at
    $0.20\,a_0^{-3}$. Periodic images are faded, and the teal box outlines the
    primitive cell; the network in \textbf{d} is a single component, so nothing
    is faded. Atom colors: white (H), gray (C), blue (N), and red (O). Bond
    cylinders indicate geometric connectivity, not bond order.}
    \label{fig:dataset}
\end{figure}

The crystals are even-electron CHNO structures, treated as closed-shell, from
AIRSS random structure searches at 1~GPa\cite{pickardAIRSSDataCarbon2020,pickardInitioRandomStructure2011},
a pressure inherited from the source searches and retained for the
fixed-geometry enthalpy comparisons. A geometric connectivity analysis
(Section~\ref{sec:structures}) classifies 61\% of the 62,109 retained
primitive cells as containing only finite bonded components, 10\% as chains,
6\% as sheets, 1\% as three-dimensional networks, and 22\% as mixed
(Fig.~\ref{fig:dataset}); the set extends down to single-atom primitive cells.
The elemental composition is near-uniform: 23\% C, 27\% H, 27\% N, and
24\% O of the 389,359 atoms in these cells. The all-electron PySCF PBE
references use the setup described in Section~\ref{sec:datagen}. All in-distribution validation and test cells have explicit unsmeared
ground-state references, which are used to evaluate the optimized energies
and densities. We use the final model
checkpoint after 50 training epochs.
Errors are reported per atom unless stated otherwise. Structure-ranking tests
also include the original plane-wave CASTEP results
\cite{clarkFirstPrinciplesMethods2005}.

\subsection{In-distribution variational optimization}
We evaluate the final periodic model on all 6211 test structures using
float64 density optimization with soft regularization $\epsilon=0.1$.
Table~\ref{tab:id_accuracy} separates three evaluations of the same learned
functional: its value at the superposition-of-atomic-densities (SAD)
initialization, its value at the fitted KS density, and its value after
density optimization from SAD. The fitted-KS-density result is a
single-point diagnostic of the learned functional, not the initialization used
for optimization. For density
coefficients $\mathbf{p}$ and fitted KS coefficients $\mathbf{p}_\mathrm{KS}$,
we define this error as
\begin{equation}
    \epsilon_\rho =
    \left[\int_\Omega
    \left|\rho_{\mathbf{p}}(\mathbf{r})
    -\rho_{\mathbf{p}_\mathrm{KS}}(\mathbf{r})\right|^2
    \mathrm{d}\mathbf{r}\right]^{1/2}
    =
    \left[\Delta\mathbf{p}^{\mathrm{T}}\mathbf{S}_\rho
    \Delta\mathbf{p}\right]^{1/2},
    \label{eq:density_error}
\end{equation}
where $\Delta\mathbf{p}=\mathbf{p}-\mathbf{p}_\mathrm{KS}$ and
$\mathbf{S}_\rho$ is the overlap matrix of the periodic density basis,
$[\mathbf{S}_\rho]_{\mu\nu}=\int_\Omega\bar\omega_\mu(\mathbf{r})\bar\omega_\nu(\mathbf{r})
\mathrm{d}\mathbf{r}$. This is an absolute cell-level error, without
normalization by the number of atoms or electrons, and is reported in
$a_0^{-3/2}$ with electron number treated as dimensionless.
We additionally report the $L^1$ density error normalized by the electron
count $N_\mathrm{e}$ of the cell,
\begin{equation}
    \eta_\rho =
    \frac{1}{N_\mathrm{e}}\int_\Omega
    \left|\rho_{\mathbf{p}}(\mathbf{r})
    -\rho_{\mathbf{p}_\mathrm{KS}}(\mathbf{r})\right|
    \mathrm{d}\mathbf{r},
    \label{eq:density_error_l1}
\end{equation}
which is dimensionless and makes errors comparable across cells with different
electron counts. We evaluate the integral on the Becke quadrature grid of
the Kohn--Sham reference calculation. It is the integrated absolute density difference per
electron; for charge-conserving differences, half of it is the fraction of the
charge that has to be moved. Both errors are measured relative to the fitted KS density and
exclude the error introduced by fitting the orbital-derived density into the
Gaussian basis.

\begin{table}[htbp]
    \centering\small
    \setlength{\tabcolsep}{2pt}
    \caption{Energy and density accuracy on all 6211 in-distribution test
    structures. The first three rows evaluate STRUCTURES-P at
    different densities; the notation method@density identifies the density at
    which the energy is evaluated. The eSEN MLIP predicts total energies directly
    from geometry and is trained without force supervision; it does not provide
    a density. APBE is a semilocal (GGA) kinetic-energy functional evaluated
    with PBE exchange--correlation and the same analytic electrostatics at the
    fitted KS density. The fitted KS density has zero density error
    by definition. $L^2$ density errors are in $a_0^{-3/2}$; $L^1$ density
    errors are per electron, in percent. MAE and median are the mean and
    median absolute energy errors relative to the PySCF KS total energy.}
    \label{tab:id_accuracy}

    \begin{tabular}{lrrrrr}
        \toprule
        Evaluation & \shortstack{MAE\\(mHa/cell)} & \shortstack{Median\\(mHa/cell)}
        & \shortstack{MAE\\(mHa/atom)}
        & \shortstack{Mean $L^2$\\density\\error $\bar{\epsilon_\rho}$}
        & \shortstack{Mean $L^1$\\density\\error $\bar{\eta_\rho}$ (\%)} \\
        \midrule
        STRUCTURES-P@SAD & 304 & 305 & 47.8 & 0.235 & 6.73 \\
        STRUCTURES-P@KS & 1.61 & 0.653 & 0.255 & 0 & 0 \\
        STRUCTURES-P@optimized & 1.58 & 0.627 & 0.254 & 0.0138 & 0.413 \\
        APBE@KS & 619 & 500 & 101 & 0 & 0 \\
        eSEN MLIP & 3.32 & 1.11 & 0.496 & -- & -- \\
    \bottomrule
    \end{tabular}

\end{table}

\begin{figure}[htbp]
    \centering
    \includegraphics[width=\textwidth]{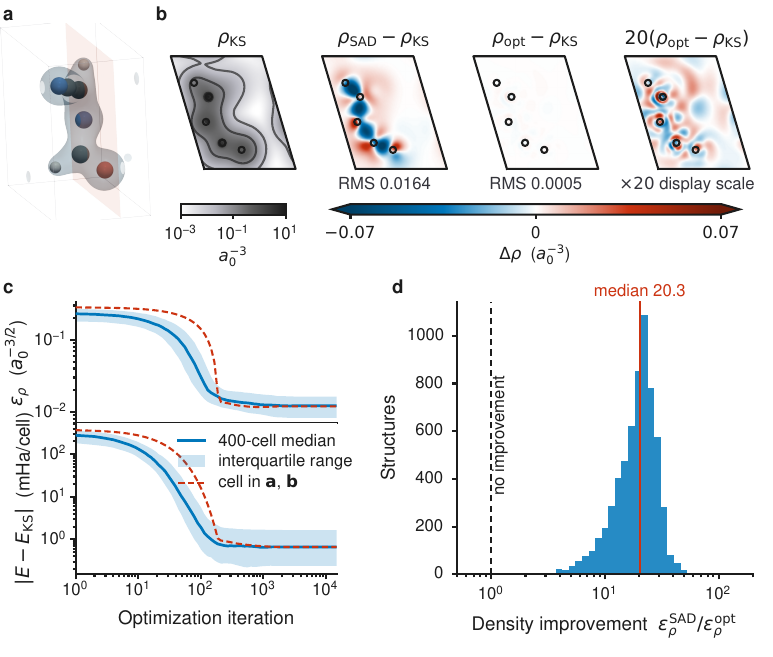}
    \caption{Density optimization reduces density and energy errors on the
    6211-cell CHNO test set. \textbf{a} Optimized density for a representative
    test cell with errors near the test-set medians (selection in
    Section~\ref{sec:density_optimization}). The molecule is centered in the cell.
    The isosurface encloses 75\% of the electrons;
    the shaded plane is used in \textbf{b}. Atom colors: white (H),
    gray (C), blue (N), and red (O). \textbf{b} Fitted Kohn--Sham density
    $\rho_\mathrm{KS}$ (logarithmic scale; contours at 0.01, 0.1, and
    $1\,a_0^{-3}$), SAD and optimized-density errors, and the latter
    magnified by 20. All error maps use a common symmetric color scale, and
    RMS denotes the root-mean-square density difference over the displayed
    plane in $a_0^{-3}$.
    Circles mark atoms within $1.1\,a_0$ of the plane.
    \textbf{c} Density and absolute energy errors during optimization:
    median and interquartile range over 400 randomly sampled test cells,
    with the representative cell overlaid. Converged runs retain their
    final values. \textbf{d} Distribution of the density-error reduction
    $\epsilon_\rho^\mathrm{SAD}/\epsilon_\rho^\mathrm{opt}$ over all 6211
    structures; dashed and red lines mark no improvement and the median,
    respectively.}
    \label{fig:variational}
\end{figure}

Table~\ref{tab:id_accuracy} shows that optimization from the SAD
initialization reduces the energy MAE by two orders of magnitude, to
1.58~mHa/cell (0.254~mHa/atom), and the mean density error by a factor of 17.
In the electron-normalized $L^1$ measure of Eq.~\eqref{eq:density_error_l1},
the mean error falls from 6.73\% to 0.413\%. The optimized energy errors have
a 90th percentile of 4.00~mHa/cell and a signed bias of $+0.252$~mHa/cell.

Figure~\ref{fig:variational} illustrates the density improvement in real space
for a test cell with energy and density errors close to the test-set medians.
On a lattice plane through this cell, the SAD
initialization misplaces charge in extended lobes between the atoms, while the
optimized-density error map is almost featureless on the same color scale
[Fig.~\ref{fig:variational}a,b]. The improvement is a property of individual
structures and not only of the average: the per-structure ratio
$\epsilon_\rho^\mathrm{SAD}/\epsilon_\rho^\mathrm{opt}$ has a median of 20.3 and
exceeds unity for all 6211 structures
[Fig.~\ref{fig:variational}d]. Note that this per-structure median is not the
same statistic as the factor 17 above, which is the ratio of the mean errors.
Most of the reduction occurs early: across the plotted sample the median density error
falls within 10\% of its final value after roughly 560 iterations
[Fig.~\ref{fig:variational}c], whereas the tighter gradient-norm criterion
below is met after a median of 1999 cycles, with a 90th percentile of 3181.

The regularized projected-gradient criterion is reached for 6206 of 6211
structures (99.9\%). The other five reach the 15000-cycle limit and are
included in all statistics. The median unregularized projected-gradient norm
at the endpoint is $1.92\times10^{-2}$ in the optimization coordinates,
compared with a median regularized norm of $1.00\times10^{-4}$.
Evaluating the learned functional at the fitted KS
density gives an energy MAE of 1.61~mHa/cell. After optimization, the learned
energy is no higher than its value at the fitted KS density for 92.8\% of
structures, and its mean change is $-0.368$~mHa/cell. The similar energy errors
at the optimized and fitted KS densities show that minimization from SAD
recovers essentially the same energy accuracy as the reference-density
diagnostic, although their signed errors can cancel differently.

As a density-independent baseline, we train the same eSEN architecture as a
machine-learned interatomic potential (MLIP) using ground-state total-energy
labels without force supervision. The density embedding and coefficient-gradient
loss are disabled. It predicts the total energy directly from the atomic geometry
and obtains an MAE of 3.32~mHa/cell (0.496~mHa/atom) on the same test set.
The baseline does not provide a density or an energy functional for trial
densities, but tests whether the variational formulation remains
competitive with direct energy regression.

At the fitted KS densities, the semilocal APBE kinetic-energy
functional\cite{constantinSemiclassicalNeutralAtom2011,constantinPerformanceSemilocalKinetic2019}
with PBE exchange and correlation gives a much larger MAE of
619~mHa/cell (101~mHa/atom; median 500~mHa) on all 6211 test
structures. This is a single-point evaluation at the fitted KS density, not an
APBE density optimization. The Wang--Teter
functional\cite{wangKineticenergyFunctionalElectron1992} could not be evaluated because its
reciprocal-space kernel exceeded the available accelerator memory.

\subsection{A unified functional for molecules and periodic crystals}
The shared density representation allows us to test whether one learned
functional can describe both isolated and periodic systems. We therefore
train a single model jointly on QM9 molecules\cite{ramakrishnanQuantumChemistryStructures2014}
and periodic CHNO crystals
and evaluate variational density optimization on held-out systems from both
domains.

The joint training set combines 110000 QM9 molecules with the 49687 periodic
CHNO training cells; the corresponding validation splits contain 10000 and
6211 systems, and the test splits contain 10831 and 6211 systems. Molecular
references use isolated all-electron PySCF PBE/6-31G(2df,p) calculations and
perturbations of the Fock matrix during SCF, rather than perturbations of a
fixed converged Fock matrix as in the periodic data. Each selected density is
one training sample, without additional domain weighting; selection details
and sample counts are given in Section~\ref{sec:training-settings}.

Both domains use the same $T_\mathrm{S}+E_\mathrm{XC}$ target and loss.
All parameters of the four-layer eSEN model, including the
element embeddings for H, C, N, O, and F, are shared. Within the network,
boundary conditions enter only through graph construction: periodic cells include neighbor edges
translated by lattice vectors, while isolated molecules contain no periodic
images.

\begin{table}[htbp]
    \centering\small
    \setlength{\tabcolsep}{4pt}
    \caption{Variational test-set accuracy of one jointly trained STRUCTURES-P
    model on
    isolated QM9 molecules and periodic CHNO crystals. All model parameters are
    shared between the two domains. Density errors are in $a_0^{-3/2}$;
    convergence refers to the regularized-gradient criterion.}
    \label{tab:joint_accuracy}

    \begin{tabular}{lrrrrr}
        \toprule
        Test domain & $N$ & \shortstack{MAE\\(mHa/cell)} & \shortstack{MAE/atom\\(mHa)} &
        \shortstack{$L^2$ density\\error} & \shortstack{Converged\\(\%)} \\
        \midrule
        QM9 & 10831 & 0.245 & 0.0146 & 0.0111 & 100.0 \\
        Periodic CHNO & 6211 & 1.77 & 0.280 & 0.0197 & 99.5 \\
    \bottomrule
    \end{tabular}

\end{table}

Table~\ref{tab:joint_accuracy} shows that the same learned functional can be
minimized successfully under both isolated and periodic boundary conditions.
The median absolute errors are 0.173~mHa/cell for QM9 and
0.727~mHa/cell for periodic CHNO.
Supplementary Note~1 compares matched small and large models
within the CHNO-only and joint training settings. For the large architecture,
the CHNO-only model reaches 1.58~mHa/cell on periodic CHNO, compared with
1.77~mHa/cell for the joint model. Joint training therefore preserves similar
periodic accuracy while extending the same functional to isolated molecules,
but does not improve the CHNO result in this comparison.

\subsection{Transferability to larger cells}
The in-distribution test contains 6211 cells with 1--9 atoms and 2--72
electrons, whereas the larger-cell test contains 734 cells with 10--24 atoms
and 22--176 electrons.
\begin{figure}[htbp]
    \centering
    \includegraphics[width=\textwidth]{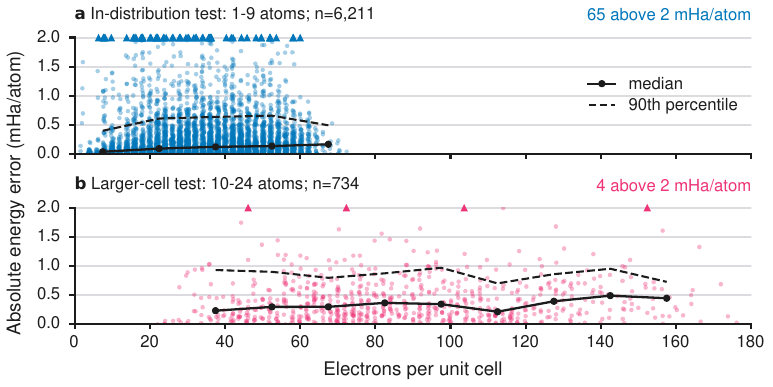}
    \caption{The mean optimized-energy error remains below 1~mHa/atom for
    cells larger than those seen in training. Each point is one structure,
    with absolute error against the Kohn--Sham reference plotted against
    electron count: \textbf{a} in-distribution and \textbf{b} larger-cell test.
    Horizontal jitter of at most 0.4 electrons separates coincident points;
    it does not affect the statistics. Solid and dashed lines show the median
    and 90th percentile in fixed 15-electron bins containing at least 12
    structures, using all errors. Both panels share the same axes. Triangles
    on the upper boundary mark errors above 2~mHa/atom, with their counts
    shown above each panel.}
    \label{fig:oodcell}
\end{figure}

Figure~\ref{fig:oodcell} compares the end-to-end error after density
optimization inside and beyond the training cell-size regime. The energy MAE
rises from 0.254 to 0.414~mHa/atom, while the 90th percentile changes from
0.613 to 0.894~mHa/atom. The median increases from 0.103 to
0.303~mHa/atom, and the mean density error from 0.0138 to 0.0307. The
cell-level signed bias is $+0.252$~mHa in distribution and $+1.121$~mHa
out of distribution. Convergence is reached for 99.9\% and 100\% of the two
sets, respectively. Thus the learned functional and variational optimization
retain sub-mHa/atom accuracy on substantially larger fixed periodic cells.

We additionally measure computational cost on 144 structures distributed over
six fixed electron-count intervals extending to 135 electrons per cell
(Section~\ref{app:timing}). All 144 calculations converge with both the full-pair
and local-pair regularizers. In the largest interval, the local-pair
approximation reduces effective end-to-end wall time from 140 to 74~s per
structure on matched hardware. Its advantage is concentrated in larger cells;
small cells show no speedup. These finite-range timings do not establish the
asymptotic scaling of the complete implementation.

\subsection{Structure rankings and thermodynamic stability}
Absolute energy errors alone do not determine usefulness for crystal-structure
prediction, which depends on energy differences within a composition and on
formation enthalpies across compositions. We assess agreement with the PySCF
training reference and include CASTEP as an additional comparison.
To match the original CASTEP dataset, we use a nominal k-point spacing of
$0.10$~\AA$^{-1}$ in PySCF. The CH$_2$ investigation shows that this sampling
is not fully converged for some structures (Supplementary Note~2).
CASTEP--PySCF differences include methodological and numerical contributions
and do not provide a controlled estimate of reference uncertainty.

\begin{figure}[!htbp]
    \centering
    \includegraphics[width=\textwidth]{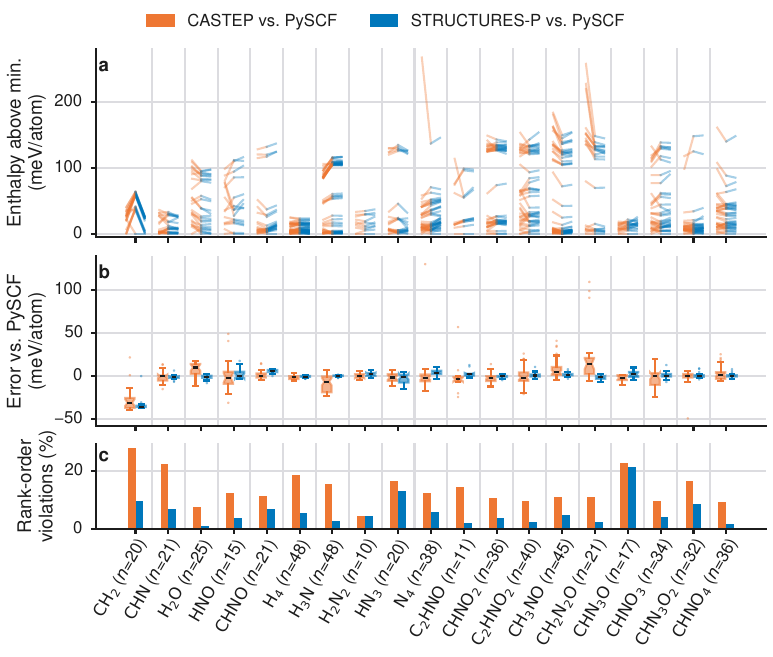}
    \caption{STRUCTURES-P reproduces the relative enthalpies of its PySCF
    training reference more closely than plane-wave CASTEP does.
    Per-composition enthalpy rankings for CASTEP, PySCF (nominal
    $0.10$~\AA$^{-1}$ k-point spacing), and
    STRUCTURES-P, using $H=E+pV$ at the $\sim$1\,GPa
    structure-search pressure.
    \textbf{a} For each composition (with $n$ structures), the enthalpy of every
    structure relative to each method's own composition minimum (meV/atom).
    Lines connect the same structure between CASTEP and PySCF (orange) and
    between PySCF and STRUCTURES-P (blue); horizontal segments indicate equal
    relative enthalpies for the two methods. \textbf{b}
    Notched box plots of relative-enthalpy deviations from PySCF (boxes:
    interquartile range; whiskers: 1.5 times that range; notches: approximate
    95\% confidence intervals for the median, which may extend beyond the
    quartiles).
    \textbf{c} Percentage of within-composition
    structure pairs whose order changes relative to PySCF. Candidates are
    selected within 150~meV/atom of the PySCF composition minimum.}
    \label{fig:rankings}
\end{figure}

The low-energy subset in Fig.~\ref{fig:rankings} contains only structures within
150~meV/atom of the PySCF minimum for its composition; neither the model nor
CASTEP predictions enter this selection.
We select candidates from the 20 most populated compositions and keep
compositions with at least ten such candidates, yielding 538 structures across
19 compositions. For each unordered
within-composition pair, the pair-energy error is the absolute difference
between the two methods' per-atom enthalpy differences, and the pair is counted
as a rank-order violation when the predicted enthalpy ordering differs from PySCF.
Formation enthalpies are calculated as
$\Delta h_\mathrm{f}=h-\sum_\alpha x_\alpha h_\alpha$, where $h$ is the structure's
enthalpy per atom, $x_\alpha$ is the atomic fraction of element $\alpha$
($\sum_\alpha x_\alpha=1$), and $h_\alpha$ is the lowest elemental enthalpy
per atom obtained by the respective method among the
evaluated structures. On the H--N edge, PySCF uses H$_8$ and N$_4$ cells,
STRUCTURES-P uses H$_6$ and N$_4$, and CASTEP uses H$_8$ and N$_8$.
These formulas describe the contents of the reference cells, not individual molecules.
We show the H--N edge because it is the most densely sampled binary slice, whereas
coverage of the full quaternary CHNO space is too sparse and uneven for a similarly
compact hull comparison.

After removing composition-dependent offsets, the CASTEP and PySCF relative
enthalpies have a Spearman correlation of 0.996, and their within-composition
pair-energy differences disagree by 35.5~meV/atom (1.30~mHa/atom) on average.
Only 3.3\% of all within-composition pairs change order, although the fraction
rises to 29.9\% for pairs separated by at most 10~meV/atom. The reference-method
disagreements are therefore concentrated among near-degenerate structures.

Figure~\ref{fig:rankings} focuses on the low-energy manifold within
150~meV/atom of the PySCF composition minimum. Across 538 structures and 19
compositions, the model changes 4.7\% of the PySCF pair rankings, compared with
13.4\% between CASTEP and PySCF. Its relative-enthalpy mean absolute error is
3.4~meV/atom, versus 8.2~meV/atom for CASTEP relative to PySCF. The model selects
the PySCF minimum in 14 of 19 compositions, while CASTEP selects it in 5 of 19.
At this level of agreement with PySCF, the numerical accuracy of the reference
data becomes an important consideration.
For CH$_2$, finer k-point sampling removes the anomalously large gap above the
original PySCF minimum. The model does not reproduce most of this artificial gap,
although it still selects the original minimum (Supplementary Fig.~1 and
Supplementary Note~2). Across the 516 screened refinement cells, refining the
spacing from 0.10 to 0.03~\AA$^{-1}$ changes the PySCF energy by 2.8~meV/atom on
average (median 0.4~meV/atom), the same order as the model's deviation from the
original reference on these cells (mean 1.7, median 1.0~meV/atom;
Supplementary Fig.~2).

\begin{figure}[htbp]
    \centering
    \centering
    \includegraphics[width=\textwidth]{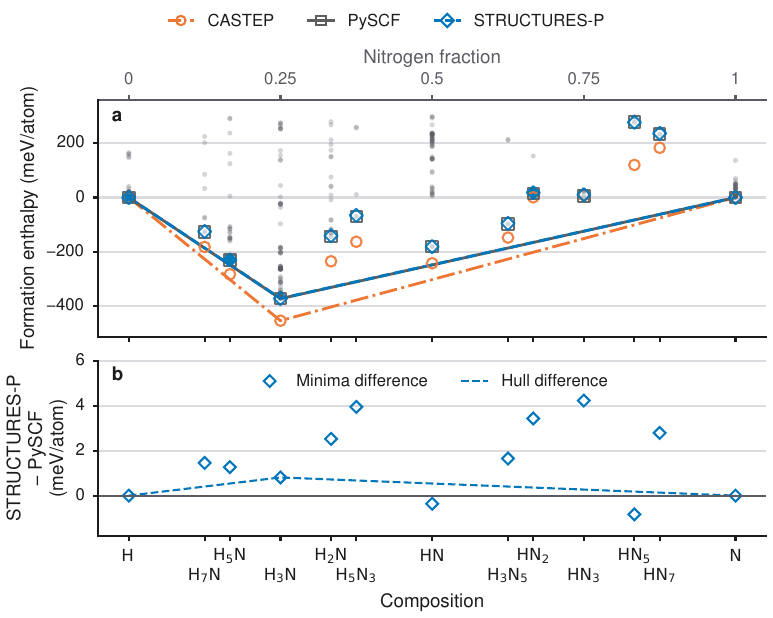}
    \caption{Formation-enthalpy convex hull along the H--N edge of the CHNO
    composition space, using $H=E+pV$ at the $\sim$1\,GPa search pressure and
    method-specific elemental references from the evaluated candidate set.
    \textbf{a} Lowest formation enthalpy at each stoichiometry and its convex
    hull for CASTEP (circles, dash-dotted), PySCF (squares, solid), and
    STRUCTURES-P (diamonds, dashed). Faint gray points show individual PySCF
    candidates with formation enthalpies up to 300~meV/atom. Filled diamonds
    mark the three stoichiometries where STRUCTURES-P and PySCF select different
    structures. All three methods give the H, H$_3$N, and N hull vertices.
    \textbf{b} STRUCTURES-P minus PySCF for the composition
    minima (diamonds) and hull (dashed line), resolved on a finer energy scale.}
    \label{fig:convexhull}
\end{figure}

The formation-enthalpy test in Fig.~\ref{fig:convexhull} extends the comparison
across compositions. On the H--N slice, the model identifies the same minimum
structure as PySCF at 10 of 13 stoichiometries and reproduces the hull vertices
at H, H$_3$N, and N. Its hull differs from the PySCF hull by 0.4~meV/atom on
average and at most 0.8~meV/atom. CASTEP gives the same hull vertices, although
its formation enthalpies differ more substantially. These hulls describe
stability within the evaluated candidate set, not a complete phase diagram.

\section{Discussion}
The central result is that a learned all-electron functional, minimized under
periodic boundary conditions from a superposition of atomic densities, recovers
Kohn--Sham energies and densities across 6211 held-out CHNO crystals, from
molecular crystals to three-dimensional networks, with a mean energy error of
0.25~mHa per atom and a mean density error of 0.4\% of the electron count.
The model represents only
$T_\mathrm{S}+E_\mathrm{XC}$, while the Hartree, electron--nuclear, and
nucleus--nucleus contributions are evaluated analytically. Unlike geometry-only
machine-learned interatomic potentials, this construction provides both an
explicit density and gradients with respect to it. The large reduction in
density error from the SAD initialization, together with the similar energy
accuracy at the optimized and fitted KS densities, indicates that the
non-ground-state training data constrain the functional along the optimization
paths relevant here.

The results test both a shared functional for molecules and crystals and
transfer to larger cells. A single jointly trained functional can be minimized for both
isolated QM9 molecules and periodic CHNO crystals, although joint training does
not improve the periodic accuracy in the matched comparison. The periodic model
retains sub-mHa/atom accuracy for cells larger than those used in training.
Likewise, the relative-energy tests show useful preservation of fixed-geometry
rankings and of the convex hull on the examined H--N slice, with remaining
differences from the PySCF reference. Because no structures are generated or
relaxed with the learned functional, these tests do not yet demonstrate a full
crystal-structure search.

The present accuracy is tied to the reference method and representation.
The reported model errors quantify agreement with the available PySCF
references rather than with uniformly converged DFT energies. The PySCF
references retain finite-orbital-basis errors for some structures. Larger
Gaussian bases can improve accuracy, but diffuse functions can cause
near-linear dependence and numerical instabilities in some periodic cells,
particularly compact high-pressure cells. The PBE approximation and density
fit introduce further limitations. The chemical scope explored here is
even-electron CHNO crystals, treated as closed-shell, near the pressure of
the AIRSS dataset, with the
larger-cell test extending system size rather than chemistry or thermodynamic
conditions.

Improving the numerical reliability of reference-data generation is an
important next step, including systematic k-point convergence checks and
more robust orbital-basis representations. All-electron methods based on
numerical atom-centered orbitals, such as FHI-aims\cite{blumAbInitioMolecular2009},
Gaussian and augmented plane waves, as in CP2K\cite{kuhneCP2KElectronicStructure2020},
or full-potential augmented plane waves, such as APW+lo in
WIEN2k\cite{blahaWIEN2k2020}, offer alternative routes to periodic reference
densities. Their use here would
still require fitting the densities into the Gaussian density representation
and constructing consistent non-ground-state energy and derivative labels.

Plane-wave pseudopotential calculations offer another route to periodic
reference data, with densities that could also be fitted into the Gaussian
density basis. However, the learned target changes: a
nonlocal pseudopotential energy depends on the one-particle density matrix and
is not directly available from an arbitrary scalar density. Incorporating this
term into a learned functional would therefore require suitable non-ground-state
energy and derivative labels, or an additional approximation to the density
matrix\cite{xuNonlocalPseudopotentialEnergy2022,maNonlocalPseudopotentialEnergy2024}.

The dense matrix operations used for global natural reparametrization and
regularization are the main scaling bottlenecks in the current implementation
(Section~\ref{app:timing}). Changes to the training loss and data generation,
together with better-conditioned density bases, could reduce the need for
these steps. Block-sparse or matrix-free implementations could also reduce
their computational cost.

Taken together, these results establish that a learned all-electron
functional can be variationally optimized in periodic systems. Broader
chemical and thermodynamic coverage and structural relaxation were outside
the scope of this work, but the present formulation does not preclude either.
Additional closed-shell chemistry and other pressures enter only through the
density basis and the training data, whereas spin-polarized systems would
require a spin-density formulation. For relaxation, the electrostatic terms and
the learned functional are differentiable with respect to atomic positions and
cell vectors. Forces and stresses at the optimized density can therefore be
obtained by automatic differentiation, up to a correction from the residual
gradient along the regularized directions. Natural next steps
are to extend the density basis and training data toward broader chemistry
and systematically converged references, to test structural relaxation, and
to replace the remaining dense regularizer operations by better-conditioned
or matrix-free schemes.
    
\section{Methods}
\subsection{Density representation in a Gaussian basis}
\label{sec:density-basis}
Following M-OFDFT\cite{zhangOvercomingBarrierOrbitalfree2024a} and
STRUCTURES25\cite{remmeStableAccurateOrbitalFree2025}, we use the atom-centered
Gaussian density expansion of Eq.~\eqref{eq:density} rather than a real-space
grid. The even-tempered basis is generated with PySCF's \texttt{aug\_etb}
from the 6-31G(2df,p) orbital basis, with exponent ratio
$\beta=2.5$\cite{sunGaussianPlanewaveMixed2017}; each function has
the usual Gaussian radial form times a spherical harmonic, with exponents
chosen as a geometric progression.

The total number of electrons in the unit cell is a linear function of the
coefficients,
\begin{equation}
    \int_\Omega \rho(\mathbf{r})\,\mathrm{d}\mathbf{r}
    =
    \mathbf{q}^{\mathrm{T}}\mathbf{p}
    =
    N_e,
    \label{eq:charge}
\end{equation}
where
\begin{equation}
    q_\mu =
    \int_\Omega
    \sum_{\mathbf{T}\in\mathcal{L}}
    \omega_\mu(\mathbf{r}-\mathbf{R}_{a(\mu)}-\mathbf{T})
    \,\mathrm{d}\mathbf{r}
    =
    \int_{\mathbb{R}^3}\omega_\mu(\mathbf{r})\,\mathrm{d}\mathbf{r}.
\end{equation}
The second equality follows because the lattice-translated unit cells tile
space. In practice, $q_\mu$ is evaluated analytically from the Gaussian basis
function. We impose this constraint during density optimization. Since
$N_e=\sum_a Z_a$, it also keeps the unit cell charge neutral, a prerequisite for
the periodic electrostatics of Section~\ref{sec:electrostatics} to be well defined.
The linear expansion does not enforce pointwise non-negativity. To quantify
violations, we sampled 50 of the 6211 test cells uniformly without replacement
and evaluated the integrated negative charge
$N_- = \int_\Omega \max[-\rho(\mathbf{r}),0]\,\mathrm{d}\mathbf{r}$ of the
optimized densities on a level-3 Becke grid; cells with detected negative
density were rechecked on level-4 and level-5 Becke grids and on
positive-weight uniform grids. The integrated negative charge was at most
$6\times10^{-5}$ electrons per cell (Supplementary Note~3).

The atom-centered basis provides a compact representation of the all-electron density,
allowing near-nuclear variations to be resolved with fewer degrees of freedom than a uniform
real-space grid. The even-tempered basis used here includes tight, high-exponent functions
centered on the nuclei.

\subsection{Periodic electrostatic energies}
\label{sec:electrostatics}
\label{app:electrostatics}
The classical electrostatic energy splits into the electron-electron (Hartree)
term $E_\mathrm{H}$, the electron-nucleus (external) term $E_\mathrm{ext}$, and
the constant nucleus-nucleus repulsion $E_\mathrm{nn}$. From here on,
$\omega_\mu$ is the lattice-periodic function $\bar\omega_\mu$ of
Section~\ref{sec:framework}, and $v_\mathrm{PBC}$ is the zero-mean periodic
Green's function of Poisson's equation, i.e., the periodic Coulomb kernel
with its zero-wave-vector component removed. The corresponding Coulomb product is
\begin{equation}
    (f \mid g)_\mathrm{PBC} =
    \int_\Omega\!\int_\Omega
    f(\mathbf{r})\,v_\mathrm{PBC}(\mathbf{r}-\mathbf{r}')\,
    g(\mathbf{r}')\,\mathrm{d}\mathbf{r}\,\mathrm{d}\mathbf{r}'.
\end{equation}

The linear density model of Eq.~(\ref{eq:density}) turns $E_\mathrm{H}$ and
$E_\mathrm{ext}$ in Eq.~\eqref{eq:model_total_energy} into a quadratic and a linear
form in the coefficients,
\begin{equation}
    E_\mathrm{H}(\mathbf{p}) = \tfrac{1}{2}\,\mathbf{p}^{\mathrm{T}}\mathbf{J}\mathbf{p},
    \qquad
    E_\mathrm{ext}(\mathbf{p}) = \mathbf{v}_\mathrm{ext}^{\mathrm{T}}\mathbf{p},
    \label{eq:estatic}
\end{equation}
with the Coulomb metric $J_{\mu\nu} = (\omega_\mu \mid \omega_\nu)_\mathrm{PBC}$
and the nuclear-attraction vector
$[\mathbf{v}_\mathrm{ext}]_\mu = -\sum_a Z_a\,(\omega_\mu \mid \delta_{\mathbf{R}_a})_\mathrm{PBC}$.

The bare periodic Coulomb sum defining $\mathbf{J}$ diverges for basis
functions with nonzero net charge; the zero-mean kernel $v_\mathrm{PBC}$
makes it well defined. We evaluate $\mathbf{J}$ and
$\mathbf{v}_\mathrm{ext}$ with the range-separated periodic Gaussian integral
scheme used by PySCF's periodic Gaussian density fitting (GDF)
machinery\cite{yeFastPeriodicGaussian2021,sunGaussianPlanewaveMixed2017,sunRecentDevelopmentsPySCF2020}. In this scheme
the Coulomb kernel is split into a short-ranged and a long-ranged part,
\begin{equation}
    \frac{1}{r} =
    \underbrace{\frac{\operatorname{erfc}(\nu r)}{r}}_{\text{short range}}
    +
    \underbrace{\frac{\operatorname{erf}(\nu r)}{r}}_{\text{long range}},
    \label{eq:rangesep}
\end{equation}
with a splitting parameter $\nu$. The short-range contribution is summed over a
finite set of real-space lattice images, while the smooth long-range contribution
is evaluated as a sum over reciprocal-lattice vectors $\mathbf{G}$ using
analytic Gaussian Fourier transforms.
For the two-center Coulomb metric in the density basis this can be written as
\begin{equation}
    J_{\mu\nu} =
    J^{\mathrm{SR}}_{\mu\nu}
    -
    \frac{\pi}{\nu^2|\Omega|}q_\mu q_\nu
    +
    \frac{4\pi}{|\Omega|}
    \sum_{\mathbf{G}\neq 0}
    \frac{\exp(-|\mathbf{G}|^2/4\nu^2)}{|\mathbf{G}|^2}\,
    \tilde{\omega}_\mu(\mathbf{G})^*
    \tilde{\omega}_\nu(\mathbf{G}),
    \label{eq:app_metric}
\end{equation}
where $J^{\mathrm{SR}}_{\mu\nu}$ is the lattice sum of short-range Gaussian
two-center integrals. The second term removes its zero-wave-vector component,
and $\tilde{\omega}_\mu(\mathbf{G})$ is the Fourier coefficient of the
periodically repeated density basis function. The $\mathbf{G}=0$ mode is omitted
consistently from the electron-electron, electron-nucleus, and nucleus-nucleus
terms; for neutral cells this omitted component cancels in the total
electrostatic energy. In the implementation, $\mathbf{v}_\mathrm{ext}$ is
obtained from the corresponding cross metric between the density basis and
PySCF's all-electron nuclear charge representation, while $E_\mathrm{nn}$ is
PySCF's point-charge Ewald energy.

Because the orbital-free density of Eq.~(\ref{eq:density}) is
lattice-periodic, no sampling of the Brillouin zone is required for any of the
electrostatic quantities.

Crucially, $\mathbf{J}$ and $\mathbf{v}_\mathrm{ext}$ depend only on the geometry
$\mathcal{M}$, not on $\mathbf{p}$. They are computed once per structure and
reused throughout density optimization, where the electrostatic energy and its
gradient reduce to
\begin{equation}
    \nabla_\mathbf{p} E_\mathrm{H} = \mathbf{J}\mathbf{p},
    \qquad
    \nabla_\mathbf{p} E_\mathrm{ext} = \mathbf{v}_\mathrm{ext}.
\end{equation}

The linear density expansion requires two-center Coulomb integrals rather than
the four-index integrals of an explicit orbital-product formulation. The Hartree
energy is a density functional in KS-DFT as well; the benefit here is that its
fixed-geometry energy and gradient evaluations reduce to matrix-vector and dot
products after precomputation.

\subsection{Data generation and density fitting}
\label{sec:datagen}
Each training example consists of a crystal geometry $\mathcal{M}$, a density
coefficient vector $\mathbf{p}$, and energy and gradient labels for the learned
part of the functional.
The model represents
$F_\theta(\mathbf{p},\mathcal{M})\approx T_\mathrm{S}[\rho]+E_\mathrm{XC}[\rho]$,
while $E_\mathrm{H}$, $E_\mathrm{ext}$, and $E_\mathrm{nn}$ are evaluated
analytically (Section~\ref{sec:electrostatics}). The label generation therefore has
three tasks: generate densities near the Kohn--Sham ground state, represent those
densities in the Gaussian density basis, and subtract the analytic
electrostatic terms evaluated at the fitted density from the Kohn--Sham energy.

\subsubsection{Structures and preprocessing}
\label{sec:structures}
We draw CHNO crystal structures from AIRSS searches performed at
1~GPa\cite{pickardAIRSSDataCarbon2020,pickardInitioRandomStructure2011}.
Compression at this pressure reduces the relative importance of dispersion
interactions\cite{conwayFirstPrinciplesCrystal2023}.
This pressure is inherited from the source searches and is also used for the
fixed-geometry enthalpy comparisons.
Before Kohn--Sham calculations, we standardize equivalent representations of the
CHNO crystal structures:
each structure is reduced to a primitive cell with
spglib\cite{togoSpglibSoftwareLibrary2024} (without idealizing the geometry),
Niggli-reduced, and given a deterministic atom ordering. We
restrict this work to the even-electron subset and treat every structure as
closed-shell in spin-restricted calculations.

For the main CHNO dataset, the local AIRSS archive contained 124710 structure
files. Excluding 176 empty files and five header-only files left 124529
readable structures, of which 62867 had an even electron count. Primitive-cell
reduction changed the electron-count parity of 274 of these structures;
excluding them left 62593 eligible structures. Of these, 62109 had retained
reference labels and entered the training, validation, and test splits.
Among the 484 omissions, 358 had documented SCF non-convergence, including
rejection of non-converged checkpoints. The remaining 126 structures
lacked reference labels and were not used; job timeouts were the main
reason.

The in-distribution split contains 49687 training, 6211 validation, and 6211
test cells with one to nine atoms and 2--72 electrons; structure identifiers
do not overlap between splits. The collection includes non-molecular
structures: six test cells contain a single atom in the primitive cell
(five carbon and one oxygen). The connectivity classes quoted in the
Results come from the rank-determination algorithm and
dimensionality score of Larsen et al.\cite{larsenDefinitionScoringParameter2019}
as implemented in ASE\cite{hjorthlarsenAtomicSimulationEnvironmenta2017},
with periodic images included and bonds defined by
$d_{ij}<k\,(r_i+r_j)$ for covalent radii $r$; each cell receives the
highest-scoring dimensionality type over the scanned interval of $k$; mixed
denotes coexisting components of different dimensionalities. The labels are
cutoff sensitive: 19.2\% of the cells change class between fixed $k=1.1$,
$1.2$, and $1.3$, so the fractions describe geometric connectivity rather
than chemically validated bonding, and the 0D class is not a molecular-crystal
count. The independent larger-cell test set is excluded from these counts.

The out-of-distribution set comprises 734 independently generated CHNO
crystals with 10--24 atoms and 22--176 electrons per cell. These candidates
were generated from random compositions and initial geometries and optimized
in separate AIRSS searches, resulting predominantly in $P1$ structures.
Because the training cells contain at most nine atoms, the two sets are
disjoint by construction. Geometry optimization used a different
pseudopotential set from the original AIRSS dataset, but all reference
energies and densities were recalculated at the fixed geometries using the
same all-electron PySCF PBE setup. The pseudopotential difference may shift the
sampled volume distribution; this test probes transfer to both larger cells
and independently generated structures, rather than a controlled size-only
extrapolation.

\subsubsection{Reference Kohn--Sham calculations}
For each crystal we first converge an all-electron, spin-restricted periodic
Kohn--Sham calculation using the PBE
functional\cite{perdewGeneralizedGradientApproximation1996} and the
6-31G(2df,p) orbital basis in PySCF\cite{sunRecentDevelopmentsPySCF2020}.
The $k$-point mesh is chosen using a reciprocal-space spacing of
$0.10$~\AA$^{-1}$, with an SCF energy tolerance of $10^{-9}$~Ha. We kept this
spacing for the full dataset to match the CASTEP reference data and because
finer k-point sampling makes the reference calculations considerably more
expensive.
The periodic reference calculations first use unsmeared occupations;
unsuccessful cases are retried with Fermi smearing of 0.001, 0.002, and
0.005~Ha. The converged calculation gives the one-particle density matrices
in the orbital basis,
\begin{equation}
    \Gamma_{\alpha\beta}^{\mathbf{k}}
    = \sum_i C_{\alpha i}^{\mathbf{k}}\,
      f_i^{\mathbf{k}}\,
      C_{\beta i}^{\mathbf{k} *},
    \label{eq:ks_density_matrix}
\end{equation}
where $\alpha$ and $\beta$ index the orbital basis functions, $i$ indexes the
Kohn--Sham orbitals, $C_{\alpha i}^{\mathbf{k}}$ are their expansion
coefficients at k-point $\mathbf{k}$, and $f_i^{\mathbf{k}}$ are the orbital
occupations.
The same calculations provide the Kohn--Sham energy components.

To test the sensitivity of the rankings to Brillouin-zone sampling
(Supplementary Note~2), we selected 637 fixed cells: all 538 ranking candidates, all 65 three-atom
CH$_2$ cells, and additional method-specific minima and nearby H--N
candidates. Each cell was first replayed at its original k-point mesh, which
had to reproduce the stored energy within 0.01~meV/atom, and then recomputed
with PySCF~2.12.0 at nominal spacings of 0.04 and 0.03~\AA$^{-1}$ with full
$\Gamma$-including meshes and otherwise unchanged settings (PBE/6-31G(2df,p),
Gaussian density fitting, the original quadrature grids, and SCF tolerances of
$10^{-11}$~Ha for the energy and $10^{-7}$ for the orbital gradient). Only
unsmeared, converged SCF results with an electron-count error below
$10^{-7}$ were retained; cells with an original occupation flag or a sampled gap below
0.1~eV at any setting were excluded. A composition passes the mesh-stability
criterion when every pairwise enthalpy difference changes by less than
0.5~meV/atom between 0.04 and 0.03~\AA$^{-1}$. All enthalpies retain the
original $pV$ term, and the model predictions are unchanged.

\subsubsection{Perturbed densities}
\label{sec:perturbed-densities}
Building on
potential-perturbation data augmentation in
KineticNet\cite{remmeKineticNetDeepLearning2023a}, we adopt the density-basis
perturbation construction of STRUCTURES25\cite{remmeStableAccurateOrbitalFree2025},
but apply it to the converged Kohn--Sham Fock matrix rather than during SCF
iterations. Sampling around this fixed reference decouples the perturbation
distribution from the SCF trajectory. A perturbing local potential is sampled
in the density basis,
$\delta v(\mathbf{r})=\sum_\mu \delta v_\mu\omega_\mu(\mathbf{r})$, and projected
to the orbital basis through the overlap three-center tensor,
\begin{equation}
    \Delta F_{\alpha\beta}^{\mathbf{k}}
    =
    \sum_\mu \delta v_\mu
    \int_\Omega
    \eta_{\alpha\mathbf{k}}^*(\mathbf{r})\,
    \omega_\mu(\mathbf{r})\,
    \eta_{\beta\mathbf{k}}(\mathbf{r})\,d\mathbf{r}.
    \label{eq:fock_perturbation}
\end{equation}

For each cell, we draw independent standard-normal coefficient vectors
$\mathbf{z}_i$ and set
$\delta\mathbf{v}_i=r_i\mathbf{z}_i/
\sqrt{\mathbf{z}_i^{\mathrm{T}}\mathbf{S}_\rho\mathbf{z}_i}$, with the
density-basis overlap matrix $\mathbf{S}_\rho$ of Eq.~\eqref{eq:density_error}.
The 40 amplitudes are descending midpoint
quantiles of a probability density that decreases linearly to zero over
$[0.01,5.6]\,E_\mathrm{h}a_0^{3/2}$.
They specify the potential norm $r_i=\lVert\delta v_i\rVert_{L^2(\Omega)}$,
not the resulting density change.

For each perturbation of Eq.~\eqref{eq:fock_perturbation}, we solve the
generalized eigenvalue problem
$(F_\mathrm{ref}^{\mathbf{k}}+\Delta F^{\mathbf{k}})C^{\mathbf{k}}
=S^{\mathbf{k}}C^{\mathbf{k}}\Lambda^{\mathbf{k}}$ once and form the
corresponding density matrix. Here, $F_\mathrm{ref}^{\mathbf{k}}$ and
$S^{\mathbf{k}}$ are the converged reference Fock and orbital-overlap
matrices, while $C^{\mathbf{k}}$ and the diagonal matrix
$\Lambda^{\mathbf{k}}$ contain the perturbed generalized eigenvectors and
eigenvalues, respectively. For unsmeared references, we reuse the ground-state
occupations and include the unperturbed density as an additional sample.
For smeared references, each perturbed one-particle problem is instead filled
in global zero-temperature Aufbau order across $k$-points, and the smeared
reference density is not included as a training sample. These replay
densities are not self-consistent zero-temperature ground states.
The training set contains 31672 cells with 40 perturbations plus an explicit
ground-state sample and 18015 cells with only the 40 replay samples, giving
2019152 density samples in total. All validation and test cells have 40
perturbations and an explicit unsmeared ground-state sample. The electronic
energy labels are evaluated at the sampled density matrices without an
electronic-entropy correction.

\subsubsection{Density fitting}
\label{app:density-fitting}
Each sampled $\Gamma^{\mathbf{k}}$ defines the Kohn--Sham density
\begin{equation}
    \rho_\mathrm{KS}(\mathbf{r})
    = \sum_{\mathbf{k}} w_{\mathbf{k}}\sum_{\alpha\beta}
      \Gamma_{\beta\alpha}^{\mathbf{k}}
      \eta_{\alpha\mathbf{k}}^*(\mathbf{r})
      \eta_{\beta\mathbf{k}}(\mathbf{r}),
\end{equation}
where $w_{\mathbf{k}}$ is the k-point weight. We re-express this density as
coefficients $\mathbf{p}$ in the Gaussian density basis. The fit minimizes the
residual Hartree energy,
\begin{equation}
    E_\mathrm{H}[\rho_\mathrm{OF}(\mathbf{p})-\rho_\mathrm{KS}]
    = \tfrac{1}{2}\mathbf{p}^{\mathrm{T}}\mathbf{J}\mathbf{p}
    - \mathbf{b}^{\mathrm{T}}\mathbf{p} + \mathrm{const},
    \label{eq:fit}
\end{equation}
where $b_\mu=(\omega_\mu\mid\rho_\mathrm{KS})_\mathrm{PBC}$. In the periodic
implementation, the contraction with the $k$-point density matrices is
\begin{equation}
    b_\mu =
    \sum_{\mathbf{k}} w_{\mathbf{k}}
    \sum_{\alpha\beta}
    (\omega_\mu \mid
    \eta_{\alpha\mathbf{k}}^*\eta_{\beta\mathbf{k}})_\mathrm{PBC}
    \Gamma_{\beta\alpha}^{\mathbf{k}} .
    \label{eq:app_coulomb_rhs}
\end{equation}
The mixed three-center Coulomb integrals in Eq.~(\ref{eq:app_coulomb_rhs}) use
the same periodic Coulomb convention as Eq.~(\ref{eq:app_metric}).
In the CHNO calculations we use a doubly constrained fit,
\begin{equation}
    \min_{\mathbf{p}}\;
    \tfrac{1}{2}\mathbf{p}^{\mathrm{T}}\mathbf{J}\mathbf{p}
    - \mathbf{b}^{\mathrm{T}}\mathbf{p}
    \quad
    \mathrm{s.t.}\quad
    \mathbf{q}^{\mathrm{T}}\mathbf{p}=N_e,\quad
    \mathbf{v}_\mathrm{ext}^{\mathrm{T}}\mathbf{p}
    = E_\mathrm{ext}^{\mathrm{KS}} ,
    \label{eq:double_constrained_fit}
\end{equation}
whose stationarity conditions yield the linear Karush--Kuhn--Tucker (KKT) system
\begin{align}
    \mathbf{J}\mathbf{p}
    - \lambda_N \mathbf{q}
    - \lambda_\mathrm{ext}\mathbf{v}_\mathrm{ext}
    &= \mathbf{b}, \\
    \mathbf{q}^{\mathrm{T}}\mathbf{p} &= N_e, \\
    \mathbf{v}_\mathrm{ext}^{\mathrm{T}}\mathbf{p}
    &= E_\mathrm{ext}^{\mathrm{KS}} .
\end{align}
The resulting density has the correct electron number and exactly reproduces the Kohn--Sham
nuclear-attraction energy in the same electrostatic convention used for
inference.

The saved quadrature diagnostics for the 6211 test ground states give a mean
fitting error $\|\rho_\mathrm{OF}-\rho_\mathrm{KS}\|_{L^2}$ of
$0.0317~a_0^{-3/2}$ and a mean relative $L^2$ error of $0.219\%$.
This fitting error is separate from the learned-model density errors, which
use the fitted KS density as their reference.

\subsubsection{Energy and gradient labels}
\label{app:gradient-label}
With $E_\mathrm{H}(\mathbf{p})$ and $E_\mathrm{ext}(\mathbf{p})$ known
analytically, we fix the $T_\mathrm{S}+E_\mathrm{XC}$ label by requiring the
orbital-free total energy to equal the Kohn--Sham reference,
\begin{equation}
    (T_\mathrm{S}+E_\mathrm{XC})(\mathbf{p})
    := E_\mathrm{elec}^{\mathrm{KS}}
    - E_\mathrm{H}(\mathbf{p}) - E_\mathrm{ext}(\mathbf{p}),
    \label{eq:kinxc_label}
\end{equation}
where $E_\mathrm{elec}^{\mathrm{KS}}=E_\mathrm{tot}^{\mathrm{KS}}-E_\mathrm{nn}$
is the Kohn--Sham electronic energy evaluated at the sampled density matrix,
without an electronic-entropy correction. This definition absorbs any density-fitting
residual into the learned target and guarantees
$E_\mathrm{tot}(\mathbf{p})=E_\mathrm{tot}^{\mathrm{KS}}$ after adding the
unchanged nuclear repulsion.

The gradient label is motivated by stationarity of the noninteracting system
used to generate each sample. At its ground state, the kinetic-energy
derivative is the negative of the effective one-particle potential, up to a
constant chemical potential that does not affect fixed-electron variations. For
a sample generated from the reference Fock matrix with perturbation
$\delta\mathbf{v}$, the stored gradient target for the learned model is
\begin{equation}
    \mathbf{y}_g
    := -\widetilde{\mathbf{v}}_\mathrm{eff}^{\mathrm{ref}}
    - \mathbf{S}_\rho\,\delta\mathbf{v}
    + \nabla_\mathbf{p}E_\mathrm{XC}(\mathbf{p}),
    \label{eq:kinxc_gradient_label}
\end{equation}
where $\widetilde{\mathbf{v}}_\mathrm{eff}^{\mathrm{ref}}$ is the effective
potential evaluated at the fitted source density and projected onto the
density basis, and $\mathbf{S}_\rho$ is the density-basis overlap matrix of
Eq.~\eqref{eq:density_error}.
The unperturbed ground-state sample is the special case $\delta\mathbf{v}=0$.

Here $v_\mathrm{eff}^{\mathrm{ref}}(\mathbf{r})$ is the converged
Kohn--Sham effective potential of the reference calculation, which defines
$F_\mathrm{ref}$. The perturbed one-particle problem has the effective potential
$v_\mathrm{eff}^{\mathrm{ref}}+\delta v$, so its non-interacting ground state
satisfies, up to a constant chemical potential,
\begin{equation}
    \frac{\delta T_\mathrm{S}}{\delta\rho}(\mathbf{r})
    =
    -v_\mathrm{eff}^{\mathrm{ref}}(\mathbf{r})
    -\delta v(\mathbf{r}) .
\end{equation}
For the exact functional in a complete basis, projecting this relation onto
the density basis gives
\begin{equation}
    \nabla_\mathbf{p}T_\mathrm{S}
    =
    -\mathbf{v}_\mathrm{eff}^{\mathrm{ref}}
    - \mathbf{S}_\rho\delta\mathbf{v}
    + \mu\mathbf{q},
\end{equation}
where $\mathbf{v}_\mathrm{eff}^{\mathrm{ref}}$ is the projection of
$v_\mathrm{eff}^{\mathrm{ref}}$ onto the density basis, and the last term
lies along the fixed-electron constraint direction and is not learned
separately.
The implemented target replaces this projection of the potential of the
orbital-derived reference density by the potential evaluated at the fitted
source density,
$\widetilde{\mathbf{v}}_\mathrm{eff}^{\mathrm{ref}}
=\mathbf{J}\mathbf{p}_\mathrm{ref}+\mathbf{v}_\mathrm{ext}
+\nabla_\mathbf{p}E_\mathrm{XC}(\mathbf{p}_\mathrm{ref})$.
Here $\mathbf{p}_\mathrm{ref}$ fits the source SCF density, including its
smearing when present. The kinetic-gradient target is
\begin{equation}
    \mathbf{g}_\mathrm{kin}
    := -\widetilde{\mathbf{v}}_\mathrm{eff}^{\mathrm{ref}}
    - \mathbf{S}_\rho\,\delta\mathbf{v}.
    \label{eq:kinetic_gradient_label}
\end{equation}
Adding the fitted-basis XC gradient
$\nabla_\mathbf{p}E_\mathrm{XC}(\mathbf{p})$ gives the stored target in
Eq.~\eqref{eq:kinxc_gradient_label}.
This finite-basis approximation does not make the stored gradient an exact
derivative of the residual energy label in Eq.~\eqref{eq:kinxc_label}.
For an explicit unperturbed ground-state sample, $\delta\mathbf{v}=0$ and
$\mathbf{p}=\mathbf{p}_\mathrm{ref}$, so the total gradient vanishes by
construction. This is not an independent check of energy--gradient consistency
after density fitting. Furthermore, reusing unsmeared
source occupations does not enforce a new global Aufbau filling after each
perturbation, so the continuum ground-state relation alone does not establish
the consistency of every finite-basis sample.

\subsection{Learned functional and training}
\label{sec:model}
\subsubsection{Coefficient transformation}
\label{app:coefficient-transform}
Training and density optimization use the global symmetric natural
reparametrization based on the density-basis overlap matrix $\mathbf{S}_\rho$.
Writing $\mathbf{S}_\rho=U_S\operatorname{diag}(s_i)U_S^{\mathrm{T}}$, the natural
transformation is
$T_\mathrm{nat}=U_S\operatorname{diag}(\sqrt{s_i})U_S^{\mathrm{T}}$.
Its inverse uses $1/\sqrt{s_i}$ directly, without eigenvalue clipping,
shifting, or truncation. A fixed orthogonal transformation $O$ then converts
the spherical-harmonic convention from PySCF to e3nn, giving the full
transformation $T=OT_\mathrm{nat}$. Subscript $0$ denotes the original
Gaussian-basis representation. The coefficients and dual charge vector
transform as $\mathbf{p}=T\mathbf{p}_0$ and
$\mathbf{q}=T^{-\mathrm{T}}\mathbf{q}_0$,
which preserves $\mathbf{q}^{\mathrm{T}}\mathbf{p}
=\mathbf{q}_0^{\mathrm{T}}\mathbf{p}_0=N_e$.
Gradients and the Coulomb matrix transform as
$\mathbf{g}=T^{-\mathrm{T}}\mathbf{g}_0$ and
$\mathbf{J}=T^{-\mathrm{T}}\mathbf{J}_0T^{-1}$, respectively; the external
potential vector follows the same dual transformation as the gradient.

\subsubsection{Functional and training objective}
The neural network represents only the kinetic and exchange--correlation part
of the orbital-free functional. For a density coefficient vector $\mathbf{p}$ and
crystal structure $\mathcal{M}$, the model predicts
\begin{equation}
    F_\theta(\mathbf{p},\mathcal{M})
    \approx T_\mathrm{S}[\rho_\mathbf{p}]
    + E_\mathrm{XC}[\rho_\mathbf{p}],
\end{equation}
with the total energy assembled as in Eq.~\eqref{eq:model_total_energy}.

We use a periodic version of the eSEN equivariant graph neural network\cite{fuLearningSmoothExpressive2025b}. The
graph contains the atoms of the reference unit cell as nodes, with edges to
neighboring atoms and their periodic images inside a fixed cutoff. Messages from
periodic images are accumulated on the corresponding atom in the reference cell,
so the model outputs one energy contribution for the unit cell. The Gaussian
density coefficients are embedded as equivariant atom-centered features using
the angular structure of the density basis. We use the linear-reference and
dimension-wise rescaling modules introduced in
M-OFDFT\cite{zhangOvercomingBarrierOrbitalfree2024a} and adopted in
STRUCTURES25\cite{remmeStableAccurateOrbitalFree2025}.
The linear reference is fitted to training energies and coefficient gradients,
capturing their dominant average contributions so that the network learns only
the residuals. Dimension-wise rescaling balances the density-feature and target
channels.

The model is trained on both energy and functional-gradient labels generated in
Section~\ref{sec:datagen}. For each training density we minimize
\begin{equation}
    \mathcal{L}
    =
    w_E
    \left|
        F_\theta(\mathbf{p},\mathcal{M}) - y_E
    \right|
    +
    w_g
    \frac{1}{N_\mathrm{OF}}
    \left\|
        \Pi_N\left[
        \nabla_\mathbf{p}F_\theta(\mathbf{p},\mathcal{M})
        - \mathbf{y}_g\right]
    \right\|_1 ,
    \label{eq:training_loss}
\end{equation}
where $y_E$ is the residual energy label of Eq.~\eqref{eq:kinxc_label} and
$\mathbf{y}_g$ is the potential-based gradient target of
Eq.~\eqref{eq:kinxc_gradient_label}.
Here $\Pi_N=I-\mathbf{q}\mathbf{q}^{\mathrm{T}}/(\mathbf{q}^{\mathrm{T}}\mathbf{q})$
projects onto variations that conserve electron number, in the transformed
coefficient representation used for training and optimization
(Section~\ref{app:coefficient-transform}). The projection removes the undetermined
chemical-potential contribution along $\mathbf{q}$.
In a batch, energy errors are averaged over systems and projected-gradient
errors over all density coefficients. The gradient term supervises the slope
of the learned energy landscape around each training density, not only its
value, so that the derivatives used during inference, obtained by
differentiating Eq.~\eqref{eq:model_total_energy} with respect to
$\mathbf{p}$, are learned directly.

The CHNO models use four eSEN layers with a $6\,a_0$ cutoff and are trained
with AdamW and a cosine learning-rate schedule. We use $\ell_1$ losses with energy
and projected-gradient weights of 0.1 and 0.9, respectively.

\subsubsection{Training settings and joint sampling}
\label{sec:training-settings}
Table~\ref{tab:training-settings} lists the training settings for the large
CHNO model and the changes used for joint QM9--CHNO training.
The joint model uses float32 matrix multiplications and the same energy and
projected-gradient loss weights as the CHNO-only model. A smaller model with
32 sphere, 32 edge, and 64 hidden channels (approximately 2.73 instead of
22.75 million parameters), trained with otherwise identical settings, is
compared with the large model in Supplementary Note~1.

For joint training, QM9 contributes the 21 perturbed SCF iterates stored at
label indices 6--26 and the converged ground state for each molecule; label
index 0 stores the initial guess and is not selected. CHNO contributes all
stored density samples, as described in Section~\ref{sec:perturbed-densities}.
The resulting training set contains 2420000 molecular and 2019152 periodic
density samples. No additional domain weight is applied, so their counts
determine the relative sampling frequency.

\begin{table}[htbp]
    \centering\small
    \setlength{\tabcolsep}{4pt}
    \caption{Training settings for the large CHNO model. The joint QM9--CHNO
    model uses the same architecture and optimizer, with a batch size of 128
    and 25 epochs.}
    \label{tab:training-settings}

    \begin{tabular}{ll}
        \toprule
        Setting & Value \\
        \midrule
        eSEN layers; angular limits $l_\mathrm{max},m_\mathrm{max}$ & 4; 4, 4 \\
        Sphere, edge, and hidden channels & 128 each \\
        Radial basis functions; neighbor cutoff & 128; $6\,a_0$ \\
        Maximum neighbors per atom & 300 \\
        Optimizer & AdamW \\
        Initial learning rate; schedule & $10^{-3}$; cosine decay to zero \\
        AdamW $\beta_1,\beta_2$; weight decay & 0.95, 0.99; $10^{-10}$ \\
        Energy and gradient loss weights $w_E,w_g$ & 0.1, 0.9 \\
        Batch size; training epochs & 64; 50 \\
        Density representation & Global symmetric natural reparametrization \\
        Model / energy-target precision & float32 / float64 \\
        Random seed & 0 \\
    \bottomrule
    \end{tabular}

\end{table}

\subsection{Density optimization}
\label{sec:density_optimization}
To solve the fixed-electron-number problem of
Eq.~\eqref{eq:density_optimization_problem}, we optimize the density coefficients
following M-OFDFT\cite{zhangOvercomingBarrierOrbitalfree2024a} and
STRUCTURES25\cite{remmeStableAccurateOrbitalFree2025}, with periodic
electrostatics evaluated analytically here.
The total gradient is
\begin{equation}
    \mathbf{g}
    =
    \nabla_\mathbf{p} F_\theta(\mathbf{p},\mathcal{M})
    + \mathbf{J}\mathbf{p}
    + \mathbf{v}_\mathrm{ext},
    \label{eq:density_optimization_gradient}
\end{equation}
where the first term is obtained by automatic differentiation of the learned
functional and the remaining terms are the analytic electrostatic gradients from
Section~\ref{sec:electrostatics}. The nuclear--nuclear repulsion is independent of
$\mathbf{p}$ and therefore does not contribute to the gradient.

To preserve the electron number, we project gradients onto the tangent space of
the constraint in Eq.~\eqref{eq:density_optimization_problem},
\begin{equation}
    \Pi_N \mathbf{g}
    =
    \mathbf{g}
    - \mathbf{q}\,
      \frac{\mathbf{q}^{\mathrm{T}}\mathbf{g}}
           {\mathbf{q}^{\mathrm{T}}\mathbf{q}} .
    \label{eq:charge_projection}
\end{equation}

All coefficient vectors, gradients, and electrostatic matrices used in the
optimization are expressed in the transformed coordinates of
Section~\ref{app:coefficient-transform}.

Some directions in the coefficient space of the orbital-free density basis are
poorly represented in the orbital-product space spanned by the Kohn--Sham basis
used to generate the training data. The learned functional is therefore only
weakly constrained along these directions, which can destabilize density
optimization. To identify them, we construct, once per structure,
\begin{equation}
    B^{\Gamma}_{\mu\alpha\beta}
    =
    \int_\Omega
    \eta_{\alpha\Gamma}^*(\mathbf{r})\,
    \omega_\mu(\mathbf{r})\,
    \eta_{\beta\Gamma}(\mathbf{r})\,d\mathbf{r},
    \qquad
    A_{\mu\nu}
    =
    \sum_{(\alpha,\beta)\in\mathcal{P}_{r_c}}
    B^{\Gamma}_{\mu\alpha\beta}B^{\Gamma}_{\nu\alpha\beta}.
    \label{eq:overlap_projection_metric}
\end{equation}
Here, $B^\Gamma$ is the three-center overlap tensor between density-basis
functions and orbital products, and $A$ is its Gram matrix over the density-basis
indices. The functions $\eta_{\alpha\Gamma}$ are $\Gamma$-point orbital basis
functions; the regularization matrix uses only their products.
For periodic cells, $\mathcal{P}_{r_c}$ includes ordered orbital-basis pairs whose
atomic centers have a minimum-image separation at most $r_c=2.5$~\AA{},
including all same-atom pairs. All density-basis functions are included.
This local-pair approximation reduces the cost of constructing $A$ with
nearly unchanged energy and density error statistics (Section~\ref{app:timing});
the molecular calculations use all orbital pairs.
In the optimized coefficient representation, the orbital-product metric becomes
$\widetilde A=T^{-\mathrm{T}}AT^{-1}$, and we write
$\widetilde A=U\Lambda U^{\mathrm{T}}$.
Its eigenvalues $\lambda_i$ measure how strongly the corresponding coefficient
directions overlap with the orbital products included in the metric.
We use the soft spectral regularizer
\begin{equation}
    R_\epsilon
    =
    (\widetilde A+\epsilon I)^{-1}\widetilde A
    =
    U\,\operatorname{diag}\!\left[
    \frac{\lambda_i}{\lambda_i+\epsilon}\right]\,
    U^{\mathrm{T}} .
    \label{eq:soft_regularization}
\end{equation}
It smoothly suppresses small-eigenvalue directions without changing the energy
evaluated at a fixed density. Without momentum, the update is
\begin{equation}
    \mathbf{p}_{t+1}
    =
    \mathbf{p}_t
    - \tau_t\,
      \Pi_N R_\epsilon\Pi_N\,
      \mathbf{g}_t ,
    \label{eq:density_optimization_update}
\end{equation}
where $\tau_t$ is the learning rate at iteration $t$.
The reported calculations use gradient descent with momentum, applied to
the full gradient of one structure; the optimizer step is
projected back onto the fixed-electron manifold, which preserves
$\mathbf{q}^{\mathrm{T}}\mathbf{p}=N_e$ also in the presence of optimizer
state. We initialize from a
superposition-of-atomic-densities guess estimated from training-set atom
statistics and normalized to the correct electron number. Unless stated
otherwise, density optimization uses $\epsilon=0.1$, momentum $0.9$,
and an adaptive learning-rate schedule initialized at $10^{-3}$ and bounded
between $10^{-3}$ and $3\times10^{-3}$. For each structure, the rate follows a
sigmoid of the log regularized-gradient norm. A rise in the moving average
of this log norm triggers decay of the upper bound toward $10^{-3}$. We monitor
both the charge-projected gradient norm $\lVert\Pi_N\mathbf{g}\rVert$ and its
regularized counterpart $\lVert R_\epsilon\Pi_N\mathbf{g}\rVert$; convergence
requires the latter to fall below $10^{-4}$. Optimization is stopped after at
most $1.5\times10^4$ cycles.
Throughout this work, numerical convergence means satisfying this
regularized-gradient criterion. Since $R_\epsilon$ suppresses weakly
constrained directions, the criterion does not establish stationarity in all
charge-conserving directions. The minimization problem in
Eq.~\eqref{eq:density_optimization_problem} defines the objective; the reported
densities are the endpoints of this regularized optimization procedure.

The representative cell in Fig.~\ref{fig:variational} was selected from the
test structures closest to the test-set medians in energy error, density
error, improvement factor, and cycle count; among these we chose a cell whose
molecule lies clear of the cell faces. It lies at the 51st, 47th, 71st, and
41st percentiles of these four quantities, respectively.

\subsection{Computational cost and size dependence}
\label{app:timing}
We report finite-size timings separately from the operation counts for
regularizer construction and application discussed below. The
benchmark contains 24 structures in each of the fixed intervals 0--20, 20--40,
40--60, 60--80, 80--100, and 100--135 electrons; the first three intervals are
drawn from the in-distribution test set and the last three from the larger-cell
test set. Density optimization uses float64 arithmetic on one NVIDIA A100 40~GB
GPU with eight host threads and batch size 1. The periodic-cell precision used
to construct the regularizer is $10^{-6}$; other optimization settings follow
Section~\ref{sec:density_optimization}. Full-pair and local-pair runs use the same
structures and run sequentially on the same GPU within each interval.
Host-side setup is prefetched and can overlap GPU optimization of another
structure, so per-structure component times need not sum to the process
throughput.
All 144 structures converge with both regularizers and in the PySCF evaluations.
The matched all-electron PySCF PBE references
use one CPU thread per calculation; the timings therefore provide practical
context rather than a hardware-normalized comparison.

\begin{figure}[htbp]
    \centering
    \includegraphics[width=\textwidth]{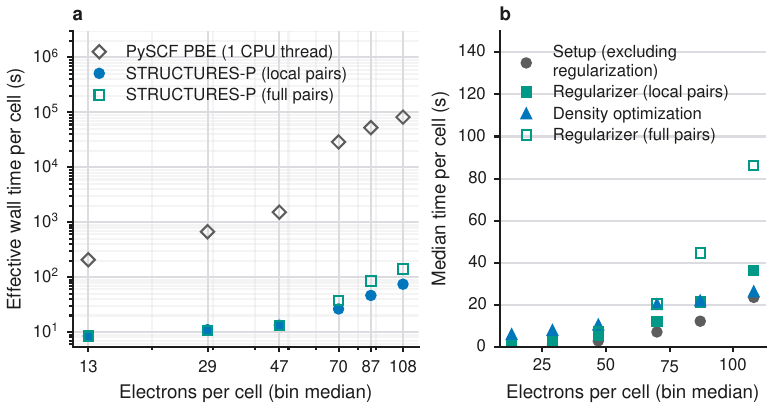}
    \caption{Computational cost over the evaluated cell-size range.
    \textbf{a} Amortized end-to-end wall time per structure for
    STRUCTURES-P with full-pair and local-pair regularization, including
    startup and compilation, and mean wall time per matched single-thread
    PySCF calculation.
    \textbf{b} Median time per structure for density optimization,
    construction of the local-pair regularizer, and the remaining static
    setup; open squares show full-pair regularizer construction.}
    \label{fig:timing}
\end{figure}

Figure~\ref{fig:timing}a covers interval medians from 13 to 108 electrons.
With local pairs, effective wall time per structure, including startup and
compilation, ranges from 8.5 to 74.5~s, compared with 8.4 to 140.1~s for full
pairs. The largest interval therefore gives a matched 1.88-fold speedup;
the three smallest intervals show no speedup. The one-thread PySCF interval
means range from 207~s to 81,800~s and provide context, not a controlled
speedup comparison. In the largest interval, median regularizer construction
falls from 86.3 to 36.4~s, whereas median density optimization changes from
25.7 to 26.3~s. The benefit thus comes from setup, not fewer optimization steps.

At fixed basis size per atom and bounded local coordination, the number of
included orbital pairs grows as $O(N)$ instead of $O(N^2)$, with $N$ the number
of atoms. The overlap factor then has $O(N^2)$ entries, and its dense Gram-matrix
construction costs $O(N^3)$ instead of $O(N^4)$. The dense change of basis and
solution for $R_\epsilon$ also cost $O(N^3)$; this is the realization used for
the reported local-pair results. Applying $\widetilde A$ through its factors and solving
$(\widetilde A+\epsilon I)\mathbf{x}=\widetilde A\mathbf{g}$ iteratively can instead cost
$O(qN^2)$ for $q$ iterations, without forming $\widetilde A$ or $R_\epsilon$.
This matrix-free alternative does not imply quadratic scaling of the complete
calculation, which still contains other dense operations, nor is a
size-independent iteration count guaranteed.

On all five periodic test evaluations (four model/training combinations on
6211 cells and the 734-cell larger-system test), replacing full pairs by local
pairs changes the energy MAE by at most 0.0051\% and the mean density error by
at most 0.076\%. The same trained functionals and reference labels are used;
only the optimization regularizer changes. Isolated-molecule evaluations and
single-point fitted-KS and geometry-only baselines are unaffected.

\subsection{Use of AI tools}
Large language model assistants (Claude, Anthropic; OpenAI Codex) were used to
write and refactor code for the project, including analysis and figure
scripts, and to draft and revise the manuscript text. All methods, results,
and conclusions were specified and verified by the authors, who take full
responsibility for the content.

\section*{Data availability}
Supplementary Data~1, which will be available with the journal version of
this article, contains the source data underlying the main-text
figures, Table~1, and the Supplementary figures, together with the lattice vectors and
atomic positions of all 62,109 CHNO cells, with their training, validation, and
test assignments, and of the 734 larger cells. The full computational dataset
(approximately 500~GB), including Kohn--Sham densities and density-functional
labels, is available from
the corresponding author on reasonable request. It has not been deposited in a
public repository because of its size and associated storage requirements.

\section*{Code availability}
The code and trained model checkpoints will be made publicly available on
GitHub upon publication.

\section*{Acknowledgements}
We warmly thank Graeme Ackland and Maurits Haverkort for very helpful discussions. This
work was supported by the Deutsche Forschungsgemeinschaft (DFG) under
Germany's Excellence Strategy EXC-2181/1 -- 390900948 (the Heidelberg
STRUCTURES Excellence Cluster), by the Wildcard Program of the
Carl-Zeiss-Stiftung, and by the Klaus Tschira Stiftung gGmbH (SIMPLAIX
program). F.A.H. and A.H. acknowledge support from the Royal Society (Grant
No.~IES/R2/242159). The authors acknowledge computing support by the state of
Baden-W\"urttemberg through bwHPC and the DFG through grant INST~35/1597-1
FUGG, and the data storage service SDS@hd supported by the Ministry of
Science, Research and the Arts Baden-W\"urttemberg (MWK) and the DFG through
grants INST~35/1803-1 FUGG and INST~35/1804-1 LAGG. Computational support
from the UK national high-performance computing service ARCHER2 is gratefully
acknowledged; access was obtained via the UKCP consortium and funded by EPSRC
grant EP/X035891/1. The funders played no role in study design, data
collection, analysis and interpretation of data, or the writing of this
manuscript.

\section*{Author contributions}
F.A.H. and A.H. conceived the project. M.K.I. developed the code, performed
all calculations and experiments, and wrote the manuscript. A.H. generated the
larger-cell test structures. F.A.H. supervised the project. All authors
discussed the results, revised the manuscript, and approved the final version.

\section*{Competing interests}
The authors declare no competing interests.

\bibliography{references}
\end{document}


\begin{center}
{\Large\bfseries Supplementary Information\par}
\vspace{7pt}
{\large All-electron machine-learned orbital-free density functional for
periodic CHNO systems\par}
\vspace{7pt}
Marc K. Ickler$^{1}$, Andreas Hermann$^{2}$, and Fred A. Hamprecht$^{1}$\par
\vspace{4pt}
{\small $^{1}$Interdisciplinary Center for Scientific Computing (IWR),
Heidelberg University, Heidelberg 69120, Germany\par
$^{2}$Centre for Science at Extreme Conditions and SUPA, School of Physics
and Astronomy, University of Edinburgh, Edinburgh EH9 3FD, United Kingdom\par}
\end{center}

\section*{Supplementary Note 1: Model-size ablation}
\label{note:model-size}
We compare two eSEN widths while holding the training data, optimization
schedule, and density-optimization protocol fixed within each training family.
The small model uses 32 sphere, 32 edge, and 64 hidden channels, whereas the
large model uses 128 channels for all three, corresponding to approximately
2.73 and 22.75 million parameters, respectively.

\begin{table}[!ht]
    \centering\small
    \setlength{\tabcolsep}{4pt}
    \footnotesize
    \caption{Model-size ablation for STRUCTURES-P on the complete test sets. The small
    and large models have approximately 2.73 and 22.75 million parameters, respectively. The
    variational density optimizations use float64 arithmetic, soft
    regularization $\epsilon=0.1$, and the same convergence threshold and cycle
    limit. Mean $L^2$ is the mean cell-level density error $\epsilon_\rho$ of
    Eq.~(4) of the main text, in $a_0^{-3/2}$; Conv.~(\%) is the percentage of structures
    meeting the regularized-gradient convergence criterion within the cycle limit
    (main-text Methods, \emph{Density optimization}). The eSEN MLIP predicts total energies
    directly from geometry and is trained without force supervision.
    Density error and convergence do not apply to this baseline.}
    \label{tab:model-size-ablation}

    \begin{tabular}{lrrrrrr}
        \toprule
        & \multicolumn{2}{c}{\shortstack{MAE\\(mHa/cell)}}
        & \multicolumn{2}{c}{Mean $L^2$}
        & \multicolumn{2}{c}{Conv. (\%)} \\
        Training setup / test & Small & Large & Small & Large & Small & Large \\
        \midrule
        STRUCTURES-P@optimized, CHNO / CHNO
            & 1.981 & 1.583 & 0.0174 & 0.0138 & 99.871 & 99.920 \\
        STRUCTURES-P@optimized, joint / QM9
            & 0.354 & 0.245 & 0.0128 & 0.0111 & 100.000 & 100.000 \\
        STRUCTURES-P@optimized, joint / CHNO
            & 2.299 & 1.771 & 0.0240 & 0.0197 & 98.905 & 99.549 \\
        \midrule
        eSEN MLIP, CHNO / CHNO
            & 3.361 & 3.316 & -- & -- & -- & -- \\
    \bottomrule
    \end{tabular}

\end{table}

Increasing the width improves both energy and density accuracy in every matched
comparison. The energy MAE decreases by 20.1\% for the CHNO-only model, 30.7\%
for joint-model QM9, and 23.0\% for joint-model CHNO; the corresponding density
error reductions are 20.7\%, 12.8\%, and 17.9\%. Convergence is unchanged or
improved for all three evaluations. In contrast, increasing the width of the
geometry-only model reduces its energy MAE by only 1.3\%, from 3.361 to
3.316~mHa/cell.
Only two widths are compared here. The larger model is therefore the best of
the tested configurations, not an established optimum; whether further
increases in capacity improve density optimization remains open.

\clearpage
\section*{Supplementary Note 2: K-point reference sensitivity}
\label{note:kpoint}
We recomputed the PySCF references of 637 fixed cells at nominal k-point
spacings of 0.04 and 0.03~\AA$^{-1}$; the cell selection, replay, refinement,
and screening protocol is given in Methods. These checks do not establish basis convergence
or the ground state across spin states. Supplementary
Figs.~\ref{fig:ch2-comparison} and~\ref{fig:shift-distribution} show the
CH$_2$ comparison and the distribution of per-cell energy changes.

\subsection*{K-point refinement of the CH$_2$ rankings}

\begin{figure}[!ht]
\centering
\includegraphics[width=\textwidth]{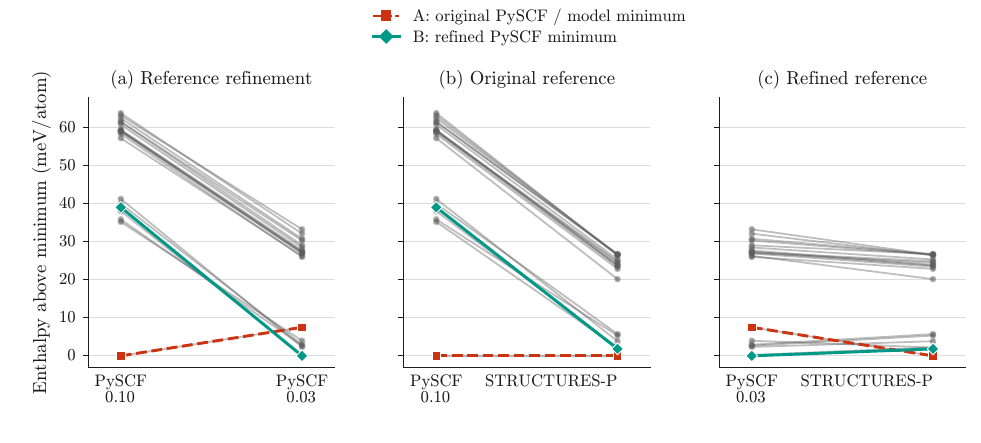}
\small
\caption{\textbf{CH$_2$ rankings before and after k-point refinement.}
Lines connect the same structures between \emph{a)} the two PySCF settings,
\emph{b)} original PySCF and STRUCTURES-P, and \emph{c)} refined PySCF and
STRUCTURES-P. Each method is referenced to its own minimum over the same
20 candidates. The highlighted structures are A, the original PySCF and model
minimum (CHNO-140055-4417-4), and B, the refined PySCF minimum
(CHNO-59219-1291-63). Spacings are in \AA$^{-1}$.}
\label{fig:ch2-comparison}
\end{figure}

Supplementary Fig.~\ref{fig:ch2-comparison} revisits the 20 CH$_2$ candidates
in the main-text ranking figure, replacing the original nominal
$0.10$~\AA$^{-1}$ PySCF references with $0.03$~\AA$^{-1}$ results.
The candidate set, geometries, $pV$ contributions, and model predictions are
unchanged. The computational protocol is described in Methods.

Finer sampling removes the isolated minimum in the original PySCF data and
changes the lowest-enthalpy candidate, while the higher- and lower-enthalpy
groups remain distinct. The model follows this broad grouping but still selects
the original minimum. For these 20 cells, the largest change in any pairwise
enthalpy difference between $0.04$ and $0.03$~\AA$^{-1}$ is
\ChTwoPairMaximum{}~meV/atom.

\subsection*{Reference energy changes and model errors}

The broader comparison uses \CommonCount{} cells with screened results at
both refined spacings, out of 637 selected candidates; \RankingCommonCount{}
belong to the main-text ranking figure. These are targeted ranking and
near-minimum candidates, not a random sample of the full dataset.

\begin{figure}[!ht]
\centering
\includegraphics[width=\textwidth]{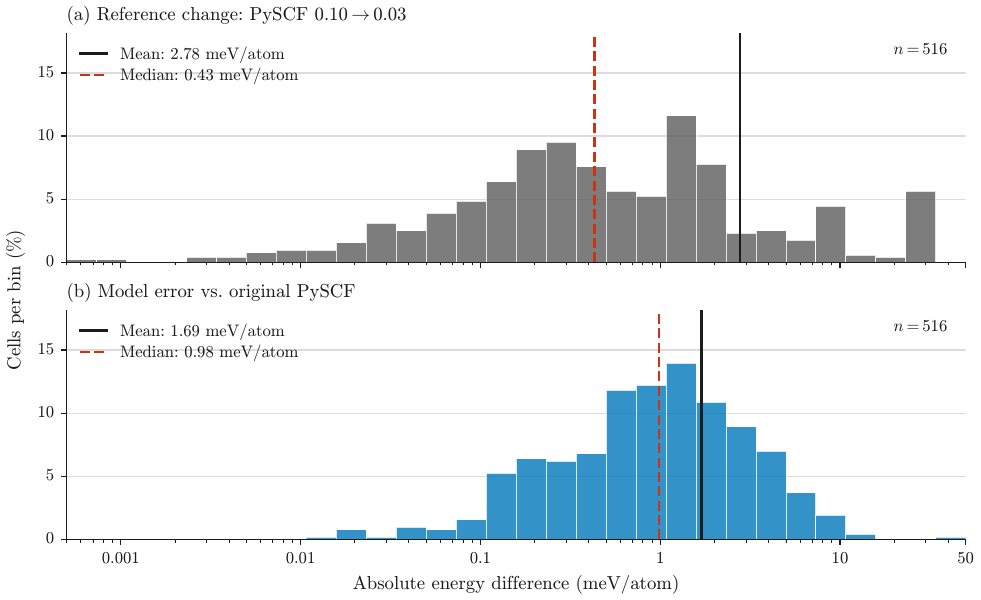}
\small
\caption{\textbf{Distributions of reference changes and model errors.}
Absolute per-atom energy differences for the same \CommonCount{} screened
cells: \emph{a)} PySCF at nominal spacings of $0.10$ and $0.03$~\AA$^{-1}$;
\emph{b)} STRUCTURES-P versus the original $0.10$~\AA$^{-1}$ PySCF reference.
Both histograms use the same 30 logarithmically spaced bins and axis scales.
Solid black and dashed red lines mark the mean and median, respectively.
At fixed geometry and $pV$, reference energy changes are also enthalpy changes.}
\label{fig:shift-distribution}
\end{figure}

Most changes are small, but a minority reach tens of meV/atom
(Supplementary Fig.~\ref{fig:shift-distribution}). The mean and median
reference changes are 2.78 and 0.43~meV/atom, compared with 1.69 and
0.98~meV/atom for the deviation of STRUCTURES-P from the original reference. The comparison does not
establish full k-point convergence: \StepLargeCount{} cells still change
by more than 5~meV/atom between $0.04$ and $0.03$~\AA$^{-1}$.

The present data support a proof of concept for learned periodic OF-DFT and
comparisons of broad energy trends. Reliable crystal-structure ranking,
especially the ordering of near-degenerate candidates, will require more
accurate reference data with systematic k-point and basis-set convergence checks.

\textbf{Source data.} The per-cell reference energies, $pV$ terms, k-point
meshes, and screening statuses of the 637 selected structures underlying
Supplementary Figs.~\ref{fig:ch2-comparison} and~\ref{fig:shift-distribution}
are included in Supplementary Data~1.

\clearpage
\section*{Supplementary Note 3: Negative-density audit}
\label{note:negative-density}
We evaluated the integrated negative charge $N_-$ of 50 randomly sampled
optimized test densities, using the protocol in Methods. All 50 optimizations
converged. On a level-3 Becke grid, 48 cells had no
detected negative density. The remaining two cases were confirmed using
level-4 and level-5 Becke grids and positive-weight uniform-grid refinements.
At level 5, CHNO-12884-189-5 had $N_-=5.99\times10^{-5}$ electrons and a
minimum density of $-2.21\times10^{-5}$ electrons/$a_0^3$, while
CHNO-84049-9284-127 had $N_-=8.40\times10^{-7}$ electrons and a minimum density
of $-1.17\times10^{-5}$ electrons/$a_0^3$. The corresponding integrated
electron counts were 53.999994 and 45.999992, compared with the target values
54 and 46, respectively. Thus the sampled optimized densities can contain
small negative regions, but their integrated negative charge is very small
relative to the unit-cell electron count.

\section*{Supplementary Data 1: Source data and cell geometries}
\label{data:source}
Supplementary Data~1 (ZIP archive, available with the journal version of this
article) contains one CSV file per figure panel or
table with the source data underlying the main-text figures, Table~1, and
Supplementary Figs.~\ref{fig:ch2-comparison} and~\ref{fig:shift-distribution},
and the lattice vectors and atomic positions (extended XYZ format, \AA) of all
62,109 CHNO cells, with their training, validation, and test assignments, and of
the 734 larger cells. A README file in the archive describes each file; cells
are identified by the same structure identifiers throughout.